\documentclass[11pt,draftclsnofoot,journal,onecolumn]{IEEEtran}

\IEEEoverridecommandlockouts

\usepackage[dvips]{graphicx}
\usepackage{subfigure}
\usepackage{amsmath}
\usepackage{amssymb}
\usepackage{verbatim}
\usepackage{url}

\newtheorem{theorem}{Theorem}

\newtheorem{observation}{Observation}

\renewcommand{\IEEEQED}{\IEEEQEDopen}

\begin{document}

\title{Performance of Joint Spectrum Sensing and MAC Algorithms for Multichannel Opportunistic Spectrum Access Ad Hoc Networks}
\author{Jihoon Park, Przemys{\l}aw Pawe{\l}czak, and Danijela \v{C}abri{\'c}%
\thanks{The authors are with the Department of Electrical Engineering, University of California, Los Angeles, 56-125B Engineering IV Building, Los Angeles, CA 90095-1594, USA (email: \{jpark, przemek, danijela\}@ee.ucla.edu).}
\thanks{Part of this work related to the performance of medium access control protocols has been presented at IEEE Symposium on New Frontiers in Dynamic Spectrum Access Networks, April 6--9, 2010, Singapore~\cite{park_dyspan_submitted}.}\thanks{Also available at http://arxiv.org/abs/0910.4704.}}

\maketitle

\begin{abstract}
We present an analytical framework to assess the link layer throughput of multichannel Opportunistic Spectrum Access (OSA) ad hoc networks. Specifically, we focus on analyzing various combinations of collaborative spectrum sensing and Medium Access Control (MAC) protocol abstractions. We decompose collaborative spectrum sensing into layers, parametrize each layer, classify existing solutions, and propose a new protocol called Truncated Time Division Multiple Access (TTDMA) that supports efficient distribution of sensing results in ``$\kappa$ out of $N$'' fusion rule. In case of multichannel MAC protocols we evaluate two main approaches of control channel design with (i) dedicated and (ii) hopping channel. We propose to augment these protocols with options of handling secondary user (SU) connections preempted by primary user (PU) by (i) connection buffering until PU departure and (ii) connection switching to a vacant PU channel. By comparing and optimizing different design combinations we show that (i) it is generally better to buffer preempted SU connections than to switch them to PU vacant channels and (ii) TTDMA is a promising design option for collaborative spectrum sensing process when $\kappa$ does not change over time.
\end{abstract}

\IEEEpeerreviewmaketitle

\section{Introduction}
\label{sec:introduction}

It is believed that Opportunistic Spectrum Access (OSA) networks will be one of the primary forces in combating spectrum scarcity~\cite{staple_spectrum_2004} in the upcoming years~\cite{Prasad_commag_2007,noam_commag_1995}. Therefore, OSA networks~\cite{Zhao_sigprocmag_2007,hoffmeyer_scc41_2008} have become the topic of rigorous investigation by the communications theory community. Specifically, the assessment of spectrum sensing overhead on OSA medium access control (MAC) performance recently gained a significant attention.

\subsection{Research Objective}
\label{sec:research_objective}

In the OSA network performance analysis, a description of the relation between the primary (spectrum) user (PU) network and the secondary (spectrum) user (SU) network can be split into two general models: macroscopic and microscopic. In the macroscopic OSA model~\cite{park_icc_2009,timmers_tvt_2009,Jia_jsac_2008} it is assumed that the time limit to detect a PU and vacate its channel is very long compared to the SU time slot, frame or packet length duration. Such a time limit is assumed to be given by a radio spectrum regulatory organization. For example, the timing requirements for signal detection of TV transmissions and low power licensed devices operating in TV bands by IEEE 802.22 networks~\cite{stevenson_commag09} (including transmission termination and channel vacancy time, i.e. a time it takes the SU to stop transmitting from the moment of detecting PU) must be equal to or smaller than 4.1\,s~\cite[Tab. 15.5]{cordeiro_book09}, while the frame and superframe duration of IEEE 802.22 are equal to 10\,ms and 160\,ms, respectively~\cite{cordeiro_book09}. Also, in the macroscopic model it is assumed that the PU channel holding time, i.e. the time in which the PU is seen by the SU as actively transmitting, is much longer than the delay incurred by the detection process performed at the SU. As a result it can be assumed in the analysis that, given high PU detection accuracy (which is a necessity), OSA network performance is determined by the traffic pattern of the SUs. That is, it depends on the total amount of data to be transmitted by the SU network, the duration of individual SU data packets and the number of SU nodes. In other words the PU bandwidth resource utilization by the SU is independent of PU detection efficiency.

In the microscopic OSA model, more popular than its macroscopic counterpart due to analytic challenges, the detection time is short in relation to the shortest transmission unit of the OSA system. Detection is also performed much more frequently than in the macroscopic model, i.e. for every SU packet~\cite{Liang_twc_2008,peh_tvt_2009} or in every time slot~\cite{jeon_twc_2008,pawelczak_tvt_2009,Papadimitratos_commag_2005,Hoang_twc_2009,wang_tmc_2009}. Also, the microscopic model assumes much higher PU activity than the macroscopic model, which justifies frequent detection cycles. Since the detection overhead is much larger than in the macroscopic model, the analysis of utilization of resources (temporarily unoccupied by PU) by OSA network cannot be decoupled from the analysis of the PU signal detection phase.

Therefore, while the distinction between macroscopic and microscopic models are somehow fluid, it is important to partition the two cases and compare them in a systematic manner. More importantly, the comparison should be based on a detailed OSA multichannel and multiuser ad hoc network model~\cite[Sec. 7.4]{hossain_book_2009}, which would not ignore the overhead from both the physical layer (PHY) and MAC layers of different cooperative and distributed spectrum sensing strategies~\cite[Tab. 7.1]{hossain_book_2009} and, in case of microscopic model, account for different channel access procedures and connection management strategies for the SUs upon PU detection, like buffering or switching to a vacant channel. Finally, the comparison should be realized using tractable analytical tools.

\subsection{Related Work}
\label{sec:related_work}

The literature on this topic can categorized into three groups: (i) performance analysis of general OSA networks, excluding a detailed model for spectrum sensing (mostly for the macroscopic model), (ii) performance of spectrum sensing isolated from MAC aspects of network collaboration, and (iii) joint performance of spectrum sensing and networking for OSA. 

One of the first works that gained insight into the general performance of OSA networks, considering impact of PU activity on blocking and throughput of the SU network was~\cite{Chou_jsac_2007}, where the capacity of a multichannel OSA system was assessed by comparing centrally coordinated versus random SU channel assignment. A spectrum sensing process was not considered. A similar problem was investigated in~\cite{srinivasa_twc_2008} where the spectrum sharing gains for PU and SU networks were obtained for a distributed and multichannel ad hoc OSA network. Unfortunately, a zero delay spectrum sensing process was assumed with genie-aided channel selection, i.e. in every time slot the receiver knew of the exact channel the transmitter will use to send data~\cite[Sec. III-C1]{srinivasa_twc_2008}.

In later works, assumptions on the OSA network model became more realistic. Specifically, Markovian analysis of SU traffic buffering on the event of PU arrival was presented for a SU exponential service time~\cite{tang_twc_2008} and for a SU phase-type service time~\cite{tang_twc_2009}. Unfortunately the impact of spectrum sensing detection time overhead on the OSA network performance was not investigated and the connection arrangement process for new SU arrivals, i.e. method to select and access a channel for a new sender-receiver pair, was assumed to be performed by a centralized entity. A different option of the above model has been analyzed in~\cite{kalil_asmta_2009}, with only PU channels dedicated to the OSA network and with a mixture of PU and SU exclusive channels. SU connection buffering was not allowed, however, SU connections were able to switch to an empty SU exclusive channel on the event of channel preemption by the PU.

A similar analysis, but with a different channelization structure, where the PU occupied more than one SU channel (contrary to~\cite{tang_twc_2008,tang_twc_2009}) was performed in~\cite{zhu_commlett_2007}. The authors addressed the cases of (i) connection blocking, and (ii) channel reservation and switching of SU connections to empty channels on PU arrival. This analysis was later extended to the case of finite SU population and packet queuing~\cite{wong_commlett_2009}, and buffering and switching of SU connections preempted by PU arrivals~\cite{zhangyan_icc_2008}. Again, in all papers listed above the spectrum sensing process was assumed to have no overhead and perfect reliability. Moreover the connection arrangement process for SUs was not considered.

A system where the PU had to wait until a SU vacates a channel was analyzed in~\cite{wong_wcnc_2008}. Both perfect and imperfect PU detection processes were considered, however detection overhead as well as a connection arrangement process for the secondary system was not considered. In~\cite{wang_twc_2009} another OSA system consisting of only two SUs was considered. Once again, SU connection arrangement process was not included in the analysis and two distinct strategies were assumed: (i) with complete SU connection blocking on the arrival of PU (just like in~\cite{kalil_asmta_2009,zhu_commlett_2007}) and (ii) with SU connection buffering on the arrival of PU (just like in~\cite{zhangyan_icc_2008}). Spectrum sensing and connection arrangement process was not investigated. A practical OSA model analyzing the impact of PU activities on the quality of Voice over Internet Protocol traffic was analyzed in~\cite{lee_commlett_2009} with a system not allowing SU connection buffering and perfect information on the PU channel presence assumed. The only work considering a microscopic model was~\cite{pawelczak_tvt_2009}, where a relation between sensing time, PU detectability and different connection arrangement processes were considered. Detailed simulations and Markov analysis were performed, however, as noted in the paper, the proposed model did not yield accurate results over the ranges of all parameters considered, e.g. level of PU activity\footnote{November 29, 2012: Refer to Section~\ref{sec:BNS} for correcting statement.}. Also, only one sensing strategy with SU connection buffering was analyzed for the case of different MAC protocols.

Considering a second group of papers (related to the performance of spectrum sensing algorithms in isolation from higher protocol layers), in~\cite{jeon_twc_2008} the analysis of the average time consumed by two stage spectrum sensing (proposed independently in~\cite{timmers_tvt_2009,wang_tmc_2009,luo_twc_2009}) was decoupled from SU traffic characteristics. Moreover the delay caused by exchanging hard decision measurements in the cooperative sensing process based on the ``and'' rule was not included. The impact of sensing overhead on, e.g., throughput, and energy consumption was explored in~\cite{hamdaoui_twc_2009}. However the analysis did not account for any OSA network and SU traffic. Also, the relationship between detection time and detection quality was not investigated. In~\cite{wang_tmc_2009} a microscopic model was analyzed with a sensing period every slot, synchronization between PU and SU, and PU stationary over the whole slot duration. Markov analysis was performed only to evaluate the delay incurred while searching for unoccupied spectrum. The sensing process was not coupled to any of the known MAC protocols and SU connections. Also, only non-collaborative spectrum sensing was considered. Finally, in~\cite{park_icc_2009}  most of the procedures related to spectrum sensing were categorized and divided into Open Systems Interconnection-like layers. Performance of the most common combinations of sensing algorithms were assessed, but only for the macroscopic model. Our analysis is the microscopic treatment of~\cite{park_icc_2009}.

Considering the final group of papers, when coupling spectrum sensing procedures with link layer protocols, there is a fundamental tradeoff between sensing time, sensing quality and OSA network throughput. This has been independently found for general OSA network models with a single sensing band~\cite{Liang_twc_2008}, multiple sensing bands~\cite{lee_twc_2008} with and without cooperative detection and centralized resource allocation, and in a context of MAC protocol abstraction~\cite{pawelczak_tvt_2009} for a non-cooperative sensing case. See also recent discussion in~\cite[Sec. 2.3.1, 7.3, and 10.2.4]{hossain_book_2009}. This tradeoff will be especially clear, while evaluating microscopic models, since the detection time creates a significant overhead for the data exchange phase. Recently the model of~\cite{Liang_twc_2008} was extended to the case of ``$\kappa$ out of $N$'' rule in cooperative sensing~\cite{peh_tvt_2009}, optimizing parameters of the model to maximize the throughput given detection rate requirements. Unfortunately, the delay caused by exchanging sensing information was not included.

\subsection{Our Contribution}
\label{sec:contribution}

In this paper, we present a unified analytical framework to design the spectrum sensing and the OSA data MAC jointly, for the macroscopic and microscopic cases. This design framework provides the (i) means of comparing different spectrum sensing techniques plus MAC architectures for OSA networks and (ii) spectrum sensing parameters such as observation time and detection rate for given design options. As a metric for optimization and comparison, we consider the average link layer OSA network throughput. Our model will account for the combined effects of the cooperative spectrum sensing and the underlying MAC protocol. For spectrum sensing, we will consider several architectures parametrized by sensing radio bandwidth, the parameters of the sensing PHY, and the parameters of the sensing MAC needed to exchange sensing data between individual OSA nodes. Along with classifying most of the well known sensing MAC protocols, we introduce a novel protocol called Truncated Time Division Multiple Access (TTDMA) that supports efficient exchange of individual sensing decisions in ``$\kappa$ out of $N$'' fusion rule. For the data MAC we will consider two protocol abstractions, (i) Dedicated Control Channel (DCC) and (ii) Hopping Control Channel (HCC), as analyzed in~\cite{pawelczak_tvt_2009,mo_tmc_2008} with novel extensions. That is, given the designs of~\cite{zhu_commlett_2007,wong_commlett_2009,zhangyan_icc_2008,lee_commlett_2009}, we will analyze MAC protocols that (i) allow (or forbid) to buffer existing SU connections on the event of PU arrival, and (ii) allow (or forbid) to switch the SU connections preempted by the PU to the empty channels. Please note that in the case of the analytical model proposed in~\cite{pawelczak_tvt_2009} for the SU connection buffering OSA MAC schemes we present an exact solution\footnote{November 29, 2012: Refer to Section~\ref{sec:BNS} for correcting statement.}. Finally, using our framework, we compute the maximum link layer throughput for most relevant combinations of spectrum sensing and MAC, optimizing parameters of the model jointly, both for the microscopic and macroscopic models.

The rest of the paper is organized as follows. System model and a formal problem description is presented in Section~\ref{sec:system_model_problem_description}. Description of spectrum sensing techniques and their analysis is presented in Section~\ref{sec:SS}. Analysis of MAC strategies are presented in Section~\ref{sec:MAC_analysis}. Numerical results for spectrum sensing process, MAC and joint design framework are presented in Section~\ref{sec:numerical_results}. Finally the conclusions are presented in Section~\ref{sec:conclusions}.

\section{System Model and Formal Problem Description}
\label{sec:system_model_problem_description}

The aim of this work is to analyze link layer throughput accounting for different combinations of MAC, spectrum sensing protocols and regulatory constraints. The model can later be used to optimize the network parameters jointly to maximize the throughput, subject to regulatory constraints. Before formalizing the problem, we need to introduce the system model, distinguishing between the microscopic and macroscopic approaches.

\subsection{System Model}
\label{sec:system_model}

\subsubsection{Microscopic Model}
\label{sec:microscopic_model}

For two multichannel MAC abstractions considered, i.e. DCC and HCC, we distinguish between the following cases: (i) when SU data transfer interrupted by the PU is being buffered (or not) for further transmission and (ii) when existing SU connection can switch (or not) to a free channel on the event of PU arrival (both for buffering and non-buffering SU connection cases). Finally, we will distinguish two cases for DCC where (i) there is a separate control channel not used by the PU and (ii) when control channel is also used by the PU for communication. All these protocols will be explained in detail in Section~\ref{sec:MAC_analysis}.

We assume slotted transmission within the SU and PU networks, where PU and SU time slots are equal and synchronized with each other. The assumptions on slotted and synchronous transmission between PU and SU are commonly made in the literature, either while analyzing theoretical aspects of OSA (see~\cite[Fig. 2]{Liang_twc_2008},~\cite[Sec. III]{pawelczak_tvt_2009},~\cite[Fig. 1]{Hoang_twc_2009},~\cite[Fig. 1]{Lai_arxiv_2007},~\cite[Fig. 5 and Sec. 5.2]{Huang_tmc_2009},~\cite[Fig. 1]{Kang_tvt_2009}) or exploring practical OSA scenarios (see~\cite[Fig. 2]{Papadimitratos_commag_2005} in the context of secondary utilization of GSM spectrum or~\cite{gronsund_pimrc_2009} in the context of secondary IEEE 802.16 resources usage). Our model can be generalized to the case where PU slots are offset in time from SU slots, however, it would require additional analysis of optimal channel access policies, see for example~\cite{Huang_tmc_2009,Gerihofer_commag_2007,Huang_infocom_2009}, which is beyond the scope of this paper. We also note that the synchrony assumption allows one to obtain upper bounds on the throughput when transmitting on a slot-asynchronous interface~\cite{gambini_twc_2008}.

\begin{figure}
\centering
\subfigure[]{\includegraphics[width=0.49\columnwidth]{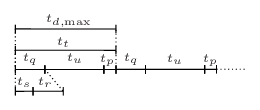}\label{fig:slot_micro}}
\subfigure[]{\includegraphics[width=0.49\columnwidth]{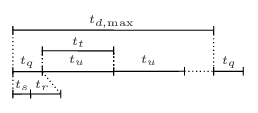}\label{fig:slot_macro}}
\caption{Difference between macroscopic and microscopic model in relation to sensing process and slot structure: (a) microscopic case and (b) macroscopic case. Symbols are explained in Section~\ref{sec:system_model} and Table~\ref{tab:parameters}.}
\label{fig:slots}
\end{figure}

The total slot duration is $t_{t}$\,$\mu$s. It is divided in three parts: (i) the detection part of length $t_{q}$\,$\mu$s, denoted as quiet time, (ii) the data part of length $t_u$\,$\mu$s, and if communication protocol requires channel switching (iii) switching part of length $t_p$\,$\mu$s. The data part of the SU time slot is long enough to execute one request to send and clear to send exchange~\cite{pawelczak_tvt_2009,mo_tmc_2008}. For the PU the entire slot of $t_{t}$\,$\mu$s is used for data transfer, see Fig.~\ref{fig:slot_micro}.

Our model assumes that there are $M$ channels having fixed capacity $C$\,Mbps that are randomly and independently occupied by the PU in each slot with probability $q_p$. There are $N$ nodes in the SU network, each one communicating directly with another SU on one of the available PU channels in one hop fashion. Also, we assume no merging of the channels, i.e only one channel can be used by a communicating pair of SUs at a time. SUs send packets with geometrically distributed length with an average of $1/q=d/(Ct_{u})$ slots for DCC, and $1/q=d/(C\left\{t_{u}+t_{p}\right\})$ slots for HCC~\cite[Sec III-C4b]{pawelczak_tvt_2009},~\cite[Sec. 3.2.3]{mo_tmc_2008}, where $d$ is the average packet size given in bits. Difference between average packet length for DCC and HCC is a result of switching time overhead for HCC, because during channel switching SUs do not transfer any data, even though they occupy the channel. We therefore virtually prolong data packet by $t_p$ for HCC to keep the comparison fair.

Every time a node tries to communicate with another node it accesses the control channel and transmits a control packet with probability $p$ to a randomly selected and non-occupied receiver. A connection is successful when only one node transmits a control packet in a particular time slot. The reason for selecting a variant of S-ALOHA as a contention resolution strategy was manyfold. First, in reality each real-life OSA multichannel MAC protocol belonging to each of the considered classes, i.e. HCC or DCC, will use its own contention resolution strategy. Implementing each and every approach in our analysis: (i) would complicate significantly the analysis, and most importantly (ii) would jeopardize the fairness of the comparison. Therefore a single protocol was needed for the analytical model. Since S-ALOHA is a widespread and well understood protocol in wireless networks and is a foundation of many other collision resolution strategies, including CSMA/CA, it has been selected for the system model herein.

In each quiet phase every SU node performs PU signal detection based on signal energy observation. Since we assume that OSA nodes are fully connected in a one hop network, thus each node observes on average the same signal realization in each time slot~\cite{peh_tvt_2009,wang_tmc_2009,visser_vtc_2008}. PU channels detected by the SU are assumed as Additive White Gaussian Noise with a channel experiencing Rayleigh fading. Therefore to increase the PU detectability by the OSA network we consider collaborative detection with hard decision combining in the detection process based on ``$\kappa$ out of $N$'' rule, as in~\cite{Zhang_ieeewc_2008,Letaief_ieeeproc_2009}. Hence we divide the quiet phase into (i) the sensing phase of length $t_s$\,$\mu$s and (ii) the reporting phase of length $t_r$\,$\mu$s. The sensing phase is of the same length for all nodes. For simplicity we do not consider in this study sensing methods that adapt the sensing time to propagation conditions as in~\cite{zhang_itw_2010}. In the sensing phase, nodes perform their local measurements. Then, during the reporting phase, nodes exchange their sensing results and make a decision individually by combining individual sensing results. We will analyze different PHY and MAC approaches to collaborative spectrum sensing, especially (i) methods to assign sensing frequencies to users, (ii) rules in combining the sensing results, and (iii) multiple access schemes for measurement reporting. In this paper we do not consider sensing strategies applicable to single channel OSA networks~\cite{jiang_twc_2009}, two stage spectrum sensing~\cite{timmers_tvt_2009}, and sensing MAC protocols based on random access~\cite{Li_ieeeglobecom_2009}, due to their excessive delay. We will explain our spectrum sensing approaches in more detail in Section~\ref{sec:SS}. Further we assume a error channel, for the sensing layer as well as for data layer where probability of error during transmission is denoted as $p_e$.

Finally, we consider two regulatory constraints under which the OSA network is allowed to utilize the PU spectrum provided the channel is idle: (i) maximum detection delay $t_{d,\max}$, i.e. a time limit within which a SU must detect a PU, and (ii) minimum detection probability $p_{d,\min}$, i.e., a probability with which a OSA system has to detect a PU signal with minimum signal to noise ratio $\gamma$. Note that in the event of mis-detection and subsequent SU transmission in a channel occupied by PU, a packet fragment is considered successfully transmitted, since in our model transmission power of SU is much higher than interference from PU, and regulatory requirements considered here do not constrain SU transmission power\footnote{The opposite case is to assume that a packet fragment is considered as lost and retransmitted. This approach however requires an acknowledgement mechanism for a lost packet fragment, see for example~\cite[Sec. II-B3]{Hoang_twc_2009},~\cite[Sec. II]{gambini_twc_2008}, that contradicts the model assumption on the geometric distribution of SU packets.} (refer for example to IEEE 802.22 draft where Urgent Coexistent Situation packets are transmitted on the same channel as active PU~\cite{stevenson_commag09,cordeiro_book09}). Moreover, maximum transmission power is a metric specific to overlay OSA systems~\cite[Sec. 2.2.5 and 8.2.1]{hossain_book_2009} where typically no spectrum sensing is considered. Also we do not consider a metric based on a maximum allowable level of collisions between PU and SU. Note that the parameters of the introduced model are summarized in Table~\ref{tab:parameters} and the abbreviations are summarized in Table~\ref{tab:abbreviations}.

\subsubsection{Macroscopic Model}
\label{sec:macroscopic_model}

We assume the same system model as for the microscopic case, except for the following differences. OSA performs detection rarely, and the PU is stable for the duration of OSA network operation, i.e. it is either transmitting constantly on a channel or stays idle. In other words quiet period occurs for multiple time slots, see Fig.~\ref{fig:slot_macro}. Also, since the PU is considered stable on every channel we do not consider all types of OSA MAC protocols introduced for the microscopic model. Instead we use classical DCC and HCC models proposed in~\cite{mo_tmc_2008} with the corrections of~\cite{pawelczak_tvt_2009} accounting for the incomplete transition probability calculations whenever OSA network occupied all PU channels and new connection was established on the control channel.

\begin{table*}
\centering
\caption{Summary of Variables and Functions Used in the Paper Grouped by Relation}
\label{tab:parameters}
\begin{tabular}{r|l|l}
\hline
Parameter & Description & Unit \\
\hline
\hline
$C$, $b$ & channel throughput, and channel bandwidth & (M)bps, (M)Hz\\
$M$, ($M_D$), $\alpha$, $N$ & total (effective) number of channels, fraction of $M$, and number of users & ---\\
$R_t$, $R$& throughput: total, excluding sensing overhead & (M)bps\\
$q_p$ $p_c$ & PU activity level, and PU activity seen by SU & ---\\
$1/q$, $d$ & average packet size & slot, bit\\
$p$ & SU channel access probability & ---\\
$t_t$, $t_u$, $t_p$, $t_q$, $t_s$, $t_r$ & length: total slot, slot data, switching time, quiet time, sensing, reporting time& ($\mu$)s\\
$t_e$, $t_a$ & time to sense, and transmit sensing information by one node & ($\mu$)s\\
$t_{d}$, $t_{d,\max}$ & detection delay, and maximum detection delay & ($\mu$)s\\
$\xi$ & slot overhead ratio & ---\\
$\epsilon$, $\kappa$, $\theta$, $\gamma$ &  time-bandwidth product, cooperation level, detection threshold, PU signal level & (M)Hzs, ---, dB, dB\\
$\Gamma(\cdot), \Gamma(\cdot,\cdot)$ & complete and incomplete gamma function & ---\\
$G_{\epsilon}(\theta)$ & function supporting computation of optimal $\theta$ & ---\\
$p_d$, $p_{d,i}$, $p_{11}$, $\hat{p}_{11}$ & detection probability: total, group, individual, individual with error & ---\\
$p_f$, $p_{f,i}$, $p_{10}$, $\hat{p}_{10}$ & false alarm probability: ----------------------------------------------------- & ---\\
$\hat{p}_{x}$, $p_{x}$ & supporting function to compute probability of detection and false alarm & --- \\
$p_{d,\min}$ & minimum probability of detection & --- \\
$p_e$ & probability of error during transmission  & --- \\
$m_{s,i}$, $n_{u,i}$, $n_g$ & number of: channels sensed per group, users in group, groups & ---\\
$m_{1,i}$, $m_{2,i}$, $\nu_i$ & supporting expressions to derive $\bar{m}_{r,i}$ for TTDMA & bit\\
$\bar{m}_r$, $\bar{m}_{r,i}$ & network, and group average number of bits to report & bit\\
$\bar{m}_s$ & number of sensing cycles & --- \\
$\beta$, $\delta$ & number of zeros, and number of bits until the end of report for TTDMA & bit \\
$t$, $i$, $j$ & time instance, and dummy variables & ---\\
$\mathcal{S}$, $s_m$ & state space of Markov chain, and size of $\mathcal{S}$ & ---\\
$k$, $l$, $m$ & number of: channels used by SU, used by PU, SU connections (previous slot) & ---\\
$x$, $y$, $z$ & ------------------------------------------------------------------------------  (current slot) & ---\\
$X_t$, $Y_t$, $Z_t$ &  ------------------------------------------------------------------------------ (at $t$) & ---\\
$\pi_{xyz}$, $p_{kl(m)|xy(z)}$ & steady state, and transition probability & ---\\
$\mathbf{1}_{x,y}$, $i_m$ & functions supporting $p_{kl|xy}$ & --- \\
$P_{x,y}^{(i)}$, $R_{x,y}^{(z)}$ & total PU arrival probability for no buffering, and buffering case& ---\\
$T_{k}^{(j)}$, $S_{m}^{(j)}$ & termination, and arrangement probability & ---\\
$\tilde{S}_{m}^{(1)}$, $\hat{S}_{m}^{(1)}$& functions supporting $S_{m}^{(j)}$ & ---\\
$\mathbb{N}$ & set of natural numbers & --- \\
$\mathbf{I}$, $\mathbf{U}$ & vector of active users, and free channels & --- \\
$I_{i,t}$, $\Pi_{i,t}$, $U_{j,t}$ & node $i$ index, priority of node $i$, and free channel $j$ index at $t$ & --- \\
$a, b, c$ ($d,e,f$) & users' (channel) enumerators & --- \\
$\mathbf{1}_{p}$ & indication function in channel switching & --- \\
$1/q_{i}$, $1/q_{t_{1}}$ ($1/q_{t_{2}}$) & SU packet length $i$, and at time $t_1$ ($t_2$) & slot \\
\hline
\end{tabular}
\end{table*}
\begin{table}
\centering
\caption{Summary of Abbreviations Used in the Paper Listed Alphabetically}
\label{tab:abbreviations}
\begin{tabular}{r|l}
\hline
Abbreviation & Explanation\\
\hline\hline
DCC & Dedicated Control Channel \\
HCC & Hopping Control Channel\\
MAC & Medium Access Control \\
OSA & Opportunistic Spectrum Access\\
PHY & Physical Layer\\
PU & Primary User\\
SPCC & Split Phase Control Channel\\
SSMA & Single Slot Multiple Access\\
SU & Secondary User\\
TDMA & Time Division Multiple Access\\
TTDMA & Truncated TDMA\\
\hline
\end{tabular}
\end{table}

\subsection{Formal Problem Description}
\label{sec:optimization_problem}

To compute the maximum throughput for different combinations of protocols and models, we define an optimization problem. The objective is the OSA network link layer throughput $R_t$. Therefore, considering the regulatory constraints given above we need to
\begin{equation}
\text{maximize } R_t=\xi R \text{ subject to } p_{d} = p_{d,\min}, t_{d} \leq t_{d,\max},
\label{eq:framework}
\end{equation}
where $t_d$ is the detection time, i.e. the time to process whole detection operation as described in Section~\ref{sec:CLP}, $R$ is the steady state link layer throughput without sensing and switching overhead, which will be computed in Section~\ref{sec:MAC_analysis}, and
\begin{equation}
\xi=
\begin{cases}
\frac{t_{t}-t_{q}-t_{p}}{t_{t}},& \text{microscopic model and DCC with channel switching},\\
\frac{t_{d,\max}-t_q}{t_{d,\max}}, & \text{macroscopic model},\\
\frac{t_{t}-t_{q}}{t_{t}},& \text{otherwise},
\end{cases}
\label{eq:xi}
\end{equation}
where $1-\xi$ is the sensing and switching overhead, see also Fig.~\ref{fig:slots}. Note that $R$ in (\ref{eq:framework}) is itself affected by $p_f$, as it will be shown in Section~\ref{sec:MAC_analysis}. Also note that $t_p$ is removed from second condition of (\ref{eq:xi}) since the switching time is negligible in comparison to inter-sensing time.

\section{Layered Model of Spectrum Sensing Analysis}
\label{sec:SS}

To design the spectrum sensing, we follow the approach of~\cite{park_icc_2009} in which the spectrum sensing process is handled jointly by (i) the sensing radio, (ii) the sensing PHY, and (iii) the sensing MAC. Using this layered model we can compare existing approaches to spectrum sensing and choose the best sensing architecture in a systematic way. Since the parameters of the design framework in (\ref{eq:framework}) are determined by the choices of individual layers, we describe and parametrize each layer of the spectrum sensing, later describing cross-layer parameters.

\subsection{Sensing Radio}
\label{sec:sensing_radio}

The sensing radio scans the PU spectrum and passes the spectrum sensing result to the sensing PHY for analysis. The sensing radio banwidth is given as $\alpha M b$, where $\alpha$ is a ratio of the bandwidth of the sensing radio to the total PU bandwidth and $b$\,MHz is the bandwidth of each PU channel\footnote{Note that $C$ used later in calculating MAC throughput is an average throughput obtained using a certain modulation over a channel with bandwidth $b$.}. With $\alpha>1/M$ node can sense multiple channels at once. However the cost of such wideband sensing radio increases. 
 
\subsection{Sensing PHY}
\label{sec:sensingPHY}

The sensing PHY analyzes the measurements from the sensing radio to determine if a PU is present in a channel. Independent of the sensing algorithm, such as energy detection, matched filter detection or feature detection~\cite{tandra_procieee_2009,yucek_commsurv_2007}, there exists a common set of parameters for the sensing PHY:  (i) time to observe the channel by one node $t_{e}$\,$\mu$s, (ii) the PU signal to noise ratio detection threshold $\theta$, and (iii) a transmit time of one bit of sensing information $t_{a}=1/C$\,$\mu$s. We denote conditional probability of sensing result $p_{ij}, i,j \in \{0,1\}$, where $j=1$ denotes PU presence and $j=0$ otherwise, and $i=1$ indicates the detection result of PU being busy and $i=0$ otherwise. Observe that $p_{10}=1-p_{00}$ and $p_{11}=1-p_{01}$.

As noted in Section~\ref{sec:system_model}, we consider energy detection as the PU detection algorithm since it does not require a priori information of the PU signal. For this detection method in Rayleigh plus Additive White Gaussian Noise channel $p_{10}$ is given as~\cite[Eq. (1)]{pawelczak_tvt_2009}
\begin{equation}
p_{10}= \frac{\Gamma (\epsilon,\theta/2)}{\Gamma(\epsilon)},
\label{eq;p10}
\end{equation}
and $p_{11}$~\cite[Eq. (3)]{pawelczak_tvt_2009}
\begin{equation}
p_{11}=e^{-\frac{\theta}{2}}\left\{ \sum_{h=0}^{\epsilon-2} \frac{\theta^h}{h!2^h} + \left(\frac{1+\gamma}{\gamma} \right)^{\epsilon-1} \left[ e^{\frac{\theta \gamma}{2+2\gamma}}- \sum_{j=0}^{\epsilon-2} \frac{(\theta \gamma)^h}{j!(2+2 \gamma)^h} \right] \right\},
\label{eq:p_support}
\end{equation}
where $\Gamma(\cdot)$ and $\Gamma(\cdot,\cdot)$ are complete and incomplete Gamma functions, respectively, and $\epsilon=\lfloor t_{e}\alpha M b\rfloor$ is a time-bandwidth product. By defining $G_{\epsilon}(\theta)=p_{10}$ and $\theta = G_{\epsilon}^{-1}(p_{10})$, we can derive $p_{11}$ as a function of $p_{10}$ and $t_{e}$.

\subsection{Sensing MAC}
\label{sec:sensing_mac}

The sensing MAC is a process responsible for sensing multiple channels, sharing sensing results with other users, and making a final decision on the PU presence. Because of the vast number of possibilities for sensing MAC algorithms it is hard to find a general set of parameters. Instead, we derive cross-layer parameters for a specific option of the sensing MAC. This methodology can be applied to any new sensing MAC scheme. We now introduce classifications which will be used in the derivation of cross-layer parameters.

\subsubsection{Sensing Strategy for Grouping Channels and Users}

Each SU has to determine which channel should be sensed among the $M$ channels. To reduce sensing and reporting overhead, OSA system can divide users and channels into $n_g$ sub-groups~\cite{biswas_icc_2009}. Sub-group $i \in \{1,\cdots,n_g\}$ is formed by $n_{u,i}$ users who should sense $m_{s,i}$ channels to make a final decision cooperatively. Assume that all users are equally divided into groups then $m_{s,i} \in \{\lfloor M/n_g \rfloor,\lceil M/n_g \rceil\}$ and $n_{u,i} \in \{\lfloor N/n_g \rfloor,\lceil N/n_g \rceil\}$. Note that for $M/n_g\in\mathbb{N}$ and $N/n_g\in\mathbb{N}$ all sub-groups have the same $n_{u,i}=N/n_g$ and $m_{s,i}=M/n_g$ for all $i$. Given $N$ and $M$, if $n_g$ is small, more users are in a group and the collaboration gain increases, but at the same time more channels must be sensed, which results in more time overhead for sensing. For large $n_g$, this relation is opposite.

\subsubsection{Combining Scheme}
\label{sec:combining_scheme}

By combining sensing results of other users, a OSA network makes a more reliable decision on PU state. As considered in~\cite{peh_tvt_2009,sun_icc_2007}, we will take $\kappa$ as a design parameter for the sensing MAC and find an optimum value to maximize the performance. Note that for the case of $N$ user cooperation if $\kappa=1$, the combining logic becomes the ``or" rule~\cite[Sec. 3.2]{hossain_book_2009},~\cite[Sec. III-C]{visser_vtc_2008} and if $\kappa=N$, it becomes the ``and'' rule.

\subsubsection{Multiple Access for Measurement Reporting}

To transmit sensing results of multiple users through the shared media, a multiple access scheme is needed. Note that this multiple access scheme is only for the reporting process, different from the multiple access for data transfer. We consider the following approaches.

\paragraph{Time Division Multiple Access (TDMA)}

This is a static and well-organized multiple access scheme for which a designated one bit slot for sensing report transmission is assigned to each user~\cite{Zhang_ieeewc_2008,biswas_icc_2009}.

\paragraph{TTDMA}

In TDMA, when the SU receives all the reporting bits from other users the SU makes a final decision of presence of PU on the channel. However, in OSA network using TTDMA SUs may not need to wait until receiving the last reporting bit, because for the ``$\kappa$ out of $N$'' rule, a reporting operation can stop as soon as $\kappa$ one bits denoting PU presence are received. This sensing MAC aims at reducing the reporting overhead, but unfortunately we have not seen any paper proposing and discussing TTDMA.

\paragraph{Single Slot Multiple Access (SSMA)}

For this scheme, known also as the boosting protocol~\cite{weiss_scvt_2003}, only one bit slot is assigned for reporting and all SUs use this slot as a common reporting period. Any SU that detects a PU transmits one bit in the common designated slot. Otherwise, a user does not transmit any bit in the designated slot. Then, reporting bits from SUs who detect a PU are overlapped and as a result all power of the slot is summed up. By measuring the power in the designated slot, a SU can determine whether the primary user exists or not. We assume perfect power control and perfect synchronization. Even though this may not be practical, because carrier frequency or the phase offset cannot be avoided in real systems, this scheme serves as an upper bound for sensing MAC performance. For the analysis of SSMA in isolation but in a more realistic physical layer conditions the reader is referred to~\cite{li_icassp_2008,zhang_twc_2009}.

\subsection{Cross-Layer Parameters}
\label{sec:CLP}

Considering the combined impact of the individual layers, we derive cross-layer parameters in the framework as described in (\ref{eq:framework}). More specifically these are $t_{q}$ and $t_d$, derived as a function of individual parameters and $p_{f}$, and $p_{d}$, denoting final network-wide probabilities of false alarm and detection, respectively.

\subsubsection{Detection Time $t_d$ and Quiet Time $t_q$}

Detection time $t_d$ is defined as the time duration from the point that a SU starts to sense, to the point that a SU makes a final decision on PU presence. Regardless of the data transfer and spectrum sensing time overlap, the final detection decision is made only after combining the sensing group's reported information~\cite{yuan_dyspan_2007}. Thus $t_d$ is the time from the start of the sensing phase to the end of the reporting phase, i.e. $t_d=t_s + t_r$. 

Since the data transfer may not be possible during sensing or reporting phases $t_q\leq t_d$, depending on the approach. When spectrum sensing and data transfer are divided in time division manner $t_{q}=t_{s}+t_{r}$. Note that three other methods sharing the same problem are possible (they will not be considered in the remainder of the paper): (i) simultaneous reporting and data, which can be implemented by using the separate channel as in~\cite{ma_dyspan_2005}, for which $t_{q}=t_{s}$, (ii) simultaneous sensing and data, implemented by using the frequency hopping method as in~\cite{wendong_commag_2007}, for which $t_{q}=t_{r}$, and (iii) simultaneous sensing, reporting, and data for which $t_{q}=0$. Conceptually, simultaneous sensing, reporting, and data transfer is possible and seems most efficient but we have not found any implementation of it in the literature. Note that in order to implement simultaneous sensing and transmission at least two radio front ends are needed, which increases the total cost of the device. 

Define $\bar{m}_s$ as the number of individual sensing events to complete sensing operation and $\bar{m}_r$ as the average number of bits to report. Then the sensing time and the reporting time can be calculated as $t_s = \bar{m}_s t_{e}$ and $t_r = \bar{m}_r t_{a}$. Note that $\bar{m}_s$ is affected by the bandwidth of the sensing radio because it can scan multiple channels at once if the bandwidth of the sensing radio is wide. For the case that the sensing radio is narrower than the bandwidth to sense, i.e. $\alpha < \max\{m_{s,1},\cdots,m_{s,n_g}\}/M$, we assume that a SU monitors all channels by sequential sensing~\cite{lee_twc_2008}, because the reporting phase should be synchronized after all SUs finish the sensing phase. With this assumption $\bar{m}_s = \left\lceil \max\{m_{s,1},\cdots,m_{s,n_g}\}/{\alpha M} \right\rceil$, because even though the bandwidth to sense is less than that of the sensing radio it still needs one sensing cycle to get information. For $\bar{m}_r$, because there are $n_g$ groups in a OSA system, $\bar{m}_r=\sum_{i=1}^{n_g}\bar{m}_{r,i}$ where $\bar{m}_{r,i}$ depends on the multiple access schemes for reporting, which we compute below.

\paragraph{TDMA}

All $n_{u,i}$ users should transmit the sensing results of $m_{s,i}$ channels. Thus, $\bar{m}_{r,i}= n_{u,i} m_{s,i}$.

\paragraph{TTDMA}

For $\kappa<n_{u,i}/2$, if $\kappa$ of ones are received, the reporting process will end. We introduce a variable $\delta$ which is the number of bits when the reporting process finishes. Thus there should be $\kappa-1$ of ones within $\delta-1$ bits and then $\delta$-th bit should be one. Because the range of $\delta$ is from $\kappa$ to $n_{u,i}$, the average number of bits for this condition is derived as 
\begin{equation}
m_{1,i}=\sum_{\delta=\kappa}^{n_{u,i}}\binom{\delta-1}{\kappa-1}\left\{(1-q_p)\delta p_{00}^{\delta-\kappa}p_{10}^\kappa + q_p \delta p_{01}^{\delta-\kappa}p_{11}^\kappa\right\}.
\label{eq:m1i}
\end{equation} 
Moreover, if the number of received zeros, denoting PU absence, equals to $n_{u,i}-\kappa+1$, the reporting process will stop because even if the remaining bits are all one, the number of ones must be less than $\kappa$. Then the reporting process stops at $\delta$-th bit if $\delta-n_{u,i}+\kappa-1$ bits of one are received within $\delta-1$ bits and zero is received at $\delta$-th bit. The range of $\delta$ is from $n_{u,i}-\kappa+1$ to $n_{u,i}$, and thus the average number of bits for this condition is
\begin{equation}
m_{2,i}=\sum_{\delta=\nu_i}^{n_{u,i}} \binom{\delta-1}{\delta-\nu_i}\left\{(1-q_p)\delta p_{00}^{\nu_i}p_{10}^{\delta-\nu_i}+q_p \delta p_{01}^{\nu_i}p_{11}^{\delta-\nu_i}\right\}, 
\label{eq:m2i}
\end{equation}
where $\nu_i=n_{u,i}-\kappa+1$. Therefore because there are $m_{s,i}$ channels to sense in a group $i$, $\bar{m}_{r,i}=m_{s,i}(m_{1,i}+m_{2,i})$.

For the case $\kappa \ge n_{u,i}/2$, $m_{1,i}$ is calculated by counting zeros and $m_{2,i}$ by counting ones. Thus, we use $\bar{m}_{r,i}=m_{s,i}(m_{1,i}+m_{2,i})$ again, by replacing $\kappa$ with $n_{u,i}-\kappa+1$, $p_{00}$ with $p_{10}$ and $p_{01}$ with $p_{11}$.

Because we assumed so far that $\kappa$ is known to each node in the network, OSA nodes know when to stop reporting measurements and start data communication without being instructed by external parties. For comparison we analyze another type of TTDMA, denoted as $\kappa$TTDMA,  where a cluster head node makes a decision to stop the reporting phase in the OSA network. For example, this approach may be necessary if the $\kappa$ value is updated in real time. In the worst case scenario this approach requires two bits to be reported by the SU, i.e. one for sending sensing data and one for an acknowledgment from the cluster head to report. Then (\ref{eq:m1i}) and (\ref{eq:m2i}) need to be modified by multiplying them by two.

\paragraph{SSMA}

For this scheme, we need only one bit per channel for reporting. Thus $\bar{m}_{r,i}=m_{s,i}$.

\subsubsection{Total False Alarm Probability $p_f$ and Detection Probability $p_d$}

Final probabilities $p_f$ and $p_d$ are obtained after cooperation, and thus affected by the sensing MAC and sensing PHY. Because each sub-group has a different number of users and channels to sense, we have 
\begin{equation}
p_f=\frac{1}{n_g} \sum_{i=1}^{n_g} p_{f,i},
\label{eq:p_f_total}
\end{equation}
where $p_{f,i}$ is the probability of false alarm of sub-group $i$. Using (\ref{eq:p_f_total}) we can also derive $p_d$ by substituting $p_{f,i}$ with $p_{d,i}$, i.e. probability of detection of sub-group $i$. The definitions of $p_{f,i}$ and $p_{d,i}$ for each protocol are as follows.

\paragraph{TDMA}

For this protocol $p_{f,i}$ is derived as
\begin{equation}
p_{f,i} = \sum_{\delta=\kappa}^{n_{u,i}} {n_{u,i} \choose \delta} \hat{p}_{10}^\delta \hat{p}_{00}^{n_{u,i}-\delta},
\label{eq:pf}
\end{equation}
where $\hat{p}_{x}=(1-p_e)p_{x}+p_e(1-p_{x})$ for $p_{x}\in\{p_{10},p_{00}\}$, while $p_{d,i}$ is derived from (\ref{eq:pf}) by substituting $\hat{p}_{10}$ with $\hat{p}_{11}$ and $\hat{p}_{00}$ with $\hat{p}_{01}$.

\paragraph{TTDMA}

In this case SU does not need to receive $n_{u,i}$ bits to make a final decision because the reporting phase is ended when the number of ones is $\kappa$. To derive $p_{f,i}$ for this case, we introduce a variable $\beta$ denoting the number of zeros. Then total number of reporting bits is $\kappa+\beta$ if the last bit is one because otherwise reporting phase will end at less than $\kappa+\beta$ bits. Therefore, there should be $\beta$ of zeros in $\kappa+\beta-1$ bits and $\kappa$-th bit should be one. Because $\beta$ can vary from 0 to $n_{u,i}-\kappa$
\begin{equation}
p_{f,i} = \sum_{\beta=0}^{n_{u,i}-\kappa} {\kappa+\beta-1 \choose \beta} \hat{p}_{10}^\kappa \hat{p}_{00}^\beta.
\label{eq:pf1}
\end{equation}
Finally $p_{d,i}$ is obtained from (\ref{eq:pf1}) by substituting $\hat{p}_{10}$ with $\hat{p}_{11}$ and $\hat{p}_{00}$ with $\hat{p}_{01}$.

\paragraph{SSMA}

Obviously, the process of the reporting information for SSMA is the same as for TDMA. Therefore $p_{f,i}$ and $p_{d,i}$ are defined the same as for TDMA.

\section{Multichannel OSA MAC Protocol Analysis}
\label{sec:MAC_analysis}

In this section we present the analysis of throughput $R$ for all considered combinations of MAC protocol architectures. As noted in Section~\ref{sec:contribution}, we propose a set of new multichannel MAC protocols for OSA. We will first describe their operation, later presenting the analysis framework.

\subsection{Description of New Multichannel MAC protocols for OSA}
\label{sec:mac_description}

We consider two major groups of MAC protocols for OSA: (i) those enabling buffering of the SU connections preempted by the PU arrival, and (ii) those enabling switching of the SU connections to a vacant channel when preempted. In the former group, when the PU arrives the existing SU connection will pause at the time of preemption and resume on the same channel as soon as the PU goes idle. We assume that the SU always waits for the PU to finish its transmission. The case where the buffered SU connection expires after a predefined time, not analyzed here, is presented in~\cite{tang_twc_2008} for the centralized network. We do not consider any channel reservation schemes for potential SU connections to be buffered~\cite{zhu_commlett_2007}. When buffering is not possible, the preempted SU connection is considered as lost and a new connection must be established on the control channel. In the latter group, when the PU arrives the existing SU connection will look for a new empty channel, to continue transmission. If such a channel cannot be found the connection is lost. Without channel switching, the exiting SU connection is lost as soon as the PU preempts the channel.

Obviously we can have four combinations of these groups for OSA MAC, which have all been considered in the analysis: (i) with no buffering and no channel switching~\cite{lee_commlett_2009} scheme denoted as B$_0$S$_0$, where SU connections preempted by PU are lost; (ii) with no buffering and channel switching~\cite{kalil_asmta_2009,zhu_commlett_2007,wong_commlett_2009} denoted as B$_0$S$_1$, where SU connections preempted by PU switch to a free channel and connections that cannot find a free channel are blocked; (iii) with buffering and no channel switching~\cite{pawelczak_tvt_2009,tang_twc_2008,tang_twc_2009} denoted as B$_1$S$_0$, where SU connections preempted by PU are being suspended from the moment of preemption until PU releases the channel; and (iv) with buffering and channel switching~\cite{zhang_icc_2008} denoted as B$_1$S$_1$, where SU connections preempted by the PU first looks for free channels, and when no free channels are found, the connections are being buffered until PU leaves the channel. The detailed procedure of distributed channel selection on the event of switching will be described in Section~\ref{sec:NBS}. Recall that the works referred above consider an OSA network with centralized channel management, providing only the upper bound on OSA network performance.

\subsection{Multichannel MAC for OSA Analysis: Preliminaries}
\label{sec:mac_preliminaries}

Usually, to compute the throughput of most non-OSA network it is assumed that a Markov chain characterizes the network state defined as the current number of connections used for data transfer~\cite[Sec. III-C]{pawelczak_tvt_2009},~\cite[Sec. 3]{mo_tmc_2008}. The state transition probability depends only on the network users' traffic characteristics. However, in the OSA system the SU data transfer connections can be terminated or delayed if a PU is detected on their channel, and thus the traffic generated by PU also affects the state transition. Moreover, with connection buffering enabled, when a PU is detected on the channel the SU does not terminate its connection but rather waits until the PU goes idle. Thus the OSA network throughput is influenced by the number of channels that are actually utilized by the SUs rather than solely by the number of SU data transfer connections.

We propose a three dimensional Markov chain of which the state vector is given as $(X_t,Y_t,Z_t)$, where $X_t, Z_t \in \mathcal{S}=\{0,1,\dots,\min(\lfloor N/2 \rfloor,M_D) \}$ and $Y_t \in \{0,1,\dots,M_D\}$, where $M_D=M-1$ for DCC and $M_D=M$ for HCC. The elements of the state vector are: (i)  $X_t$ denoting the number of channels that are actually utilized by the SUs at time $t$, (ii) $Y_t$ denoting the number of channels on which the PU is detected at time $t$, and (iii) $Z_t$ denoting the number of connections for the data transmission between the OSA users at time $t$. This distinction allows to compute the exact channel utilization, contrary to~\cite[Sec. III]{tang_twc_2008} where buffered SU connections were also considered to be utilizing the PU channels.

Considering a real OSA system, there are conditions that qualify valid states. With SU connection buffering-enabled MAC protocols for OSA, the number of connections cannot be less than the number of channels utilized by SUs, i.e. $X_t \le Z_t$. Additionally, SUs do not pause transmissions over unoccupied channels. Therefore, the number of SU connections not utilizing a channel cannot exceed the number of channels occupied by PUs, i.e. $Z_t-X_t \le Y_t$ or $Z_t \le X_t + Y_t$. Finally, the sum of the channels utilized by PUs and the SUs cannot be greater than $M_D$, i.e. $X_t+Y_t \le M_D$. By combining these conditions we can compactly write them as
\begin{equation}
0\leq X_t \le Z_t \le X_t + Y_t \le M_D.
\label{eq:Cond}
\end{equation}

When connection buffering is disabled the number of SU connections must be the same as the number of channels utilized by SUs, i.e. $X_t = Z_t$. Therefore, for non-buffering SU connection OSA MAC protocols $(X_t,Y_t,Z_t=X_t)\Rightarrow(X_t,Y_t)$.

For the microscopic case the average channel throughput, excluding switching and sensing overhead, is computed as
\begin{equation}
R =C \sum_{x=0}^{s_m} \sum_{y=0}^{M_D} \sum_{z=0}^{s_m} x \pi_{xyz},
\label{eq:R_micro}
\end{equation}
where $s_m=\max\{\mathcal{S}\}$ and the steady-state probability $\pi_{xyz}$ is given by
\begin{equation}
\pi_{xyz}=\lim_{t \rightarrow \infty} \Pr(X_{t}=x,Y_{t}=y,Z_{t}=z), 
\label{eq:pi}
\end{equation}
and the state transition probabilities to compute (\ref{eq:pi}) will be derived in the subsequent section, uniquely for each OSA multichannel MAC protocol. 

Finally, for the macroscopic case the average channel throughput, excluding switching and sensing overhead, is computed as
\begin{equation}
R=\{q_p(1-p_d)+(1-q_p)(1-p_f)\}R_cC,
\label{eq:R_macro}
\end{equation}
where $R_{c}=\sum_{i=1}^{s_m}i \pi_{i}$ and $\pi_{i}$ is a solution to a steady state Markov chain given by~\cite[Eq. (13)]{pawelczak_tvt_2009}. Since the macroscopic model assumes no PU activity in each time slot, SU connection buffering and switching is not needed. Note that contrary to the incorrect assumptions of~\cite[Eq. (12)]{pawelczak_tvt_2009},~\cite[Eq. (7) and (9)]{mo_tmc_2008} we compute $R$ in (\ref{eq:R_micro}) and (\ref{eq:R_macro}) taking all the channels into account, irrespective of the type of OSA MAC. This is because models of~\cite{pawelczak_tvt_2009,mo_tmc_2008} considered only data channels for the throughput investigation in DCC in the final calculation stage, assuming that no data traffic is being transmitted on control channel. However, the utilization must be computed over all channels, irrespective of whether one channel transmitted only control data or not.

\subsection{Derivation of State Transition Probabilities for the Microscopic Model}
\label{sec:state_transition_probabilities}

We denote the state transition probability as
\begin{equation}
p_{xyz|klm}=Pr(X_{t}=x,Y_{t}=y,Z_{t}=z|X_{t-1}=k,Y_{t-1}=l,Z_{t-1}=m).
\label{eq:pklmn}
\end{equation}
Note that changes in $X_t$ and $Z_t$ depend on the detection of the PU. In addition, changes in $Z_t$ depend on OSA traffic characteristics such as the packet generation probability $p$ and the average packet length $1/q$. Also, note that the steady state probability vector $\mathbf{\pi}$ containing all possible steady state probabilities $\pi_{xyz}$ is derived by solving $\mathbf{\pi}=\mathbf{\pi}\mathbf{P}$, where entries of right stochastic matrix $\mathbf{P}$ are defined as (\ref{eq:pklmn}) knowing that $\sum_{x,y,z}\pi_{xyz}=1$.

As a parameter to model PU state, $p_c$ denotes the probability that a OSA network collectively detects a PU channel as occupied\footnote{Note that, contrary to~\cite[Eq. (8)]{pawelczak_tvt_2009}, we do not consider packet capture effects in the definition of $p_{c}$ as the packet capture model proposed in~\cite{pawelczak_tvt_2009} was an approximation.}, i.e.
\begin{equation}
p_{c}=q_{p} p_{d}+(1-q_{p})p_{f}. 
\label{eq:P1}
\end{equation}

We introduce two supporting functions. First, we denote $T_k^{(j)}$ as the probability of termination of $j$ SU connections at time $t$ given that $k$ channels are utilized by the OSA network at time $t-1$, which is derived as~\cite[Eq. (2)]{mo_tmc_2008}
\begin{equation}
T_{k}^{(j)} = 
\begin{cases} 
\binom{k}{j} q^j (1-q)^{k-j}, & k \ge j>0,\\
0, & \text{otherwise}.
\end{cases}
\label{eq:Tkj}
\end{equation}
Note that $k$ in $T_k^{(j)}$ denotes the number of channels utilized by OSA network rather than the number of SU connections because only active connections can be terminated at the next time slot. And second, we denote $S_{m}^{(j)}$ as the probability of $j$ SU successful new connections at time $t$, given $m$ connections were active at time $t-1$. We need to modify the definition of $S_{m}^{(j)}$ given in~\cite[Eq. (5) and (8)]{mo_tmc_2008} considering PU detection on the control channel. If a PU is detected on a control channel, an SU connection cannot be generated because there is no chance to acquire a data channel. We then have~\cite[Eq. (17)]{pawelczak_tvt_2009}
\begin{equation}
S_{m}^{(j)}=
\begin{cases}
\tilde{S}_{m}^{(1)}, & j=1 \text{ (DCC)},\\
\tilde{S}_{m}^{(1)}\frac{N-2m-1}{N-1} \frac{M_D-m}{M}, & j=1 \text{ (HCC)},\\
1-S_{m}^{(1)}, & j=0,\\
0, & \text{otherwise},
\end{cases}
\label{eq:SkjDCC}
\end{equation}
where
\begin{equation}
\tilde{S}_{m}^{(1)}=
\begin{cases}
\hat{S}_{m}^{(1)}, & \text{PU free control channel, DCC only},\\
(1-p_{c})\hat{S}_{m}^{(1)}, & \text{otherwise},
\end{cases}
\end{equation}
and $\hat{S}_{m}^{(1)}= (N-2m) p (1-p)^{N-2m-1}$. Again, note that for the SU connection buffering protocols the sub-parameter $m$ of $S_{m}^{(j)}$ is not the number of channels utilized by SUs, but the number of SU connections. This is because we assume that a SU that has a connection but pauses data transmission due to the PU presence does not try to make another connection. We can now derive the transition probabilities individually for all four different OSA MAC protocols.

\subsubsection{Case B$_0$S$_0$}
\label{sec:NBNS}

Recall that for non-buffering OSA MAC protocols $Z_t=X_t$. Thus $p_{kl|xy}$ is defined as (\ref{eq:pklmn}) without $Z_{t}$. Because it is assumed that no more than one connection can be generated in one time slot, it is impossible to transit from $k$ connections at time $t-1$ to $x>k+1$ connections at time $t$. The state transition probability for this condition is 0.

For $x=k+1$ only one SU connection is created, no current connection is terminated and $y$ PUs can appear on the channels that are not utilized by SU. Thus, from $M_D-x$ channels, a PU appears on $y$ channels, so $\binom{M_D-x}{y}$ cases of PU appearances are possible.

Now, consider the case $x < k+1$. When a SU data connection is terminated, there can be two possible reasons: (i) a SU completes its transmission, or (ii) a PU is detected on a channel that is assigned to a SU for data transmission before sensing. The former was analyzed~\cite[Sec. 3]{mo_tmc_2008}. To model the latter, we introduce variable $i$ denoting the number of channels that were reserved for SU data transmission before sensing but cannot be utilized due to PU detection. We have the following observation.
\begin{observation}
For multichannel OSA MAC without SU connection buffering or channel switching the number of PU appearance combinations is $\binom{x+i}{i}\binom{M_D-x-i}{y-i}$.
\begin{proof}
When the OSA network detects PU on $i$ channels from $x+i$ channels that are going to be utilized by SUs before sensing, there can be $\binom{x+i}{i}$ possible combinations for the PU appearance on the channels. For the remaining $M_D-x-i$ channels, $y-i$ channels should be occupied by the PU because the total number of channels in which a PU is detected at time $t$ should be $y$. Thus there are $\binom{M_D-x-i}{y-i}$ possible combinations for PU appearance on unassigned channels.
\end{proof}
\label{prop:prop1}
\end{observation}

For the SU traffic generation in the case of $x<k+1$, there are two possible cases: (i) no connection is created and $k-x+i$ connections are terminated, and (ii) one connection is created and $k-x-i+1$ connections are terminated. Recall that $k$ connections at $t-1$ are changed to $x+i$ connections at $t$ before sensing. Also note that $i\in\left[0,1,\dots,\min \left(s_m-x,y\right)\right]$, where $s_m-x$ is the number of possible SU connections that can be terminated by PU appearance. By summing over all possible $i$ we can compute the transition probability for the case $x<k+1$.

In addition, we need to discuss the edge state\footnote{As shown in~\cite[Sec. III-D]{pawelczak_tvt_2009} the edge state was not considered in~\cite[Eq. (6)]{mo_tmc_2008}, which resulted in an incorrect model.} which considers two cases: (i) no more channels are available, either utilized by SUs or PUs, and (ii) all possible SU connections are established\footnote{If $s_m=M_D$ there can be many free SUs but no channel available. On the other hand, if $s_m=\lfloor N/2 \rfloor$ there can be one free SU for even $N$ or no free SU for odd $N$.} which we denote as ``full connection state''. For the transition from full connection state to edge state, we have to consider the case that one new connection is generated while any existing connection is not terminated, which means a trial for the new connection by the free SU is not established because there already exists all possible connections.

Writing all conditions compactly, denote the indicator for the edge state
\begin{equation}
\mathbf{1}_{x,y}=
\begin{cases}
1, & x+y=M_D \textrm{~or~} x=s_m,\\
0, & \textrm{otherwise},
\end{cases}
\label{eq:Ed}
\end{equation}
and define $P_{x,y}^{(i)} = {x+i \choose i}{M_D-x-i \choose y-i}p_{c}^y(1-p_{c})^{M_D-y}$, the complete state transition probability is given as
\begin{equation}
p_{xy|kl}=
\begin{cases}
0, & x > k+1\\
T_k^{(0)} S_k^{(1)} P_{x,y}^{(0)}, & x=k+1,\\
\displaystyle \sum_{i=0}^{i_{m}} \left(T_{k}^{(k-x-i)} S_k^{(0)} + T_k^{(k-x-i+1)} S_k^{(1)}\right)P_{x,y}^{(i)}, & x<k+1, k<s_m \text{~or~} \mathbf{1}_{x,y}=0,\\
\begin{split}
\hspace{-0.15em}\sum_{i=0}^{i_{m}}& \left(T_k^{(k-x-i)} S_k^{(0)} + T_k^{(k-x-i+1)} S_k^{(1)}\right)P_{x,y}^{(i)}\\
&+T_k^{(0)} S_k^{(1)} P_{0,y}^{(0)},
\end{split} & x<k+1, k=s_m, \mathbf{1}_{x,y}=1,
\end{cases}
\label{eq:p_nobuf_noswc}
\end{equation}
where $i_{m}=\min(s_m-x,y)$.

\subsubsection{Case B$_0$S$_1$}
\label{sec:NBS}

Although in the SU connection non-switching case both DCC and HCC can be considered, only DCC will be able to perform switching without any additional control data exchange, which we prove formally. 

Before going into detail of the derivation note that for the class of OSA MAC protocols with a dedicated control channel every node can follow the connection arrangement of the entire network. Because the dedicated control channel is continuously monitored by all network nodes via a separate front end~\cite[Sec. 2.2]{mo_tmc_2008}, each node can learn the overall network configuration. Note that this also applies to Split Phase Control Channel MAC (SPCC)~\cite[Sec. 2.4]{mo_tmc_2008},~\cite[Fig. 2(c)]{pawelczak_tvt_2009} as well, since SPCC has a dedicated control channel phase. For HCC, as well as Multiple Rendezvous Control Channel~\cite{mo_tmc_2008} it is impossible for a single node to learn the whole network connection arrangement since each sender receiver pair cannot listen to others while following its own hopping sequence. We now present the following proof.
\begin{theorem}
Channel switching in DCC can be performed without any additional control message exchange.
\begin{proof}
We prove this by showing a possible distributed channel switching process. Following earlier observation, in DCC each node can trace the connection arrangement of others, i.e. which channel has been reserved by a sender receiver pair. To distribute the switching events equally among SUs each SU computes the priority level as
\begin{equation}
\Pi_{i,t}=\Pi_{i,t-1}+\mathbf{1}_{p},
\label{eq:priority}
\end{equation}
where
\begin{equation}
\mathbf{1}_{p}=
\begin{cases}
1, & \text{preemption by PU},\\
0, & \text{otherwise},
\end{cases}
\end{equation}
and $\Pi_{i,t}$ is the priority level of SU $i$ at time $t$. For $\Pi_{i,0}\notin \mathbb{N}$ the priority is a MAC address of the SU, transformed into a real number for each SU by a network-wide known function. Now, having a set of priorities of all communicating node pairs, the OSA network is able to select, in a distributed manner, a new set of communication channels upon PU arrival as
\begin{equation}
\underset{{\mathbf{I}}}{\underbrace{\{I_{a,t},I_{b,t},\dots,I_{c,t}\}}}\rightarrow\underset{{\mathbf{U}}}{\underbrace{\{U_{d,t},U_{e,t},\dots,U_{f,t}\}}},
\end{equation}
where $\mathbf{|I|}=\mathbf{|U|}=M_D-X_{t}-Y_{t}$, $\rightarrow$ is the mapping operator denoting process of switching active SU connection $i$ to free channel $j$, $I_{i,t}$ denotes index of communicating SUs (transmitters) at time $t$, where $\Pi_{a,t}>\Pi_{b,t}>\dots>\Pi_{c,t}$ and $U_{j,t}$ denotes free channel with index $j$ at $t$.
\end{proof}
\label{thm:1}
\end{theorem}
Note that existing connections that have not been mapped to a channel are considered blocked. Also note that algorithm given in Theorem~\ref{thm:1} connections are preempted randomly with equal probability by PU. Since new SU connections are also assumed to use new channels randomly with equal probability, each SU connection is blocked with uniform probability.

To enable SU connection switching in HCC one way is to augment it with a separate radio front end which would follow the hopping sequences and control data exchange of the OSA network. Obviously this increases the cost of hardware and contradicts the idea of HCC, where all channels should be used for data communication. Therefore while evaluating OSA MAC protocols in Section~\ref{sec:mac_performance}, we will not consider SU connection switching for HCC.

We now define the state transition probability $p_{xy|kl}$ for the considered OSA MAC protocol. Because $x>k+1$ is infeasible, the state transition probability for $x>k+1$ equals to zero. For $x=k+1$, $y$ PUs can appear on any of $M_D$ channels because even though a PU is detected, the SUs can still transmit data by switching to the idle channels and the possible number of PU appearances is $\binom{M_D}{y}$. Note that the possible number of PU appearances in the case B$_0$S$_1$ is always $\binom{M_D}{y}$, even for the edge state, because the data channel can be changed by switching to a vacant channel after the PU detection. Because it is impossible to create more than one new connection at a time, the OSA connection creation probabilities for $x=k+1$ are the same as in (\ref{eq:p_nobuf_noswc}), i.e. $T_{k}^{(0)}S_{k}^{(1)}$.

For $x < k+1$, if SUs are not in an full connection state, there are two cases for the OSA traffic generation: (i) no connection is created and $k-x$ connections are terminated, and (ii) one connection is created and $k-x+1$ connections are terminated. On the other hand, for the state transition to the edge state we use variable $i$, just like in the case B$_0$S$_0$, to derive the probabilities because connections can be terminated by a PU in the full connection state state. Furthermore, for the transition from the full connection state to the full connection state, we should take into account $T_k^{(0)} S_k^{(1)}$ again.

With all conditions the state transition probabilities are denoted compactly as
\begin{equation}
p_{xy|kl}=
\begin{cases}
0, & x>k+1,\\
T_{k}^{(0)}S_{k}^{(1)}P_{0,y}^{(0)}, & x=k+1,\\
\left(T_{k}^{(x-k)}S_{k}^{(0)}+T_{k}^{(x-k+1)}S_{k}^{(1)}\right)P_{0,y}^{(0)}, & x < k+1, \mathbf{1}_{x,y}=0,\\
\displaystyle \sum_{i=0}^{i_{m}}\left(T_k^{(k-x-i)} S_k^{(0)} + T_k^{(k-x-i+1)} S_k^{(1)}\right)P_{0,y}^{(0)}, & x < k+1, k< s_m, \mathbf{1}_{x,y}=1,\\
\begin{split}
\hspace{-0.15em}
\sum_{i=0}^{i_{m}}&\left(T_k^{(k-x-i)} S_k^{(0)} + T_k^{(k-x-i+1)} S_k^{(1)}\right)P_{0,y}^{(0)}\\
&\qquad+T_k^{(0)} S_k^{(1)}P_{0,y}^{(0)},
\end{split} & x < k+1, k=s_m, \mathbf{1}_{x,y}=1.\\
\end{cases}
\label{eq:p_nobuf_swc}
\end{equation}

\subsubsection{Case B$_1$S$_0$}
\label{sec:BNS}

Before we discuss this case we present the following observation, which implicates the design of simulation models and derivation of $p_{xyz|klm}$ for SU connection buffering MAC protocols.
\begin{observation}
For all SU connection buffering OSA MAC protocols the same average link level throughput results from creating a brand new connection or resuming a previously preempted and buffered connection on the arrival of PU on a channel. 
\begin{proof}
Due to the memoryless property of the geometric distribution
\begin{equation}
\Pr(1/q_{i}>1/q_{t_{1}}+1/q_{t_{2}}|1/q_{i}>1/q_{t_{1}})=\Pr(1/q_{i}>1/q_{t_{2}}),
\label{eq:q_i}
\end{equation}
where $1/q_{i}$ is the duration of connection $i$, $1/q_{t_{1}}$ is the connection length until time $t_1$ when it has been preempted by PU, and $1/q_{t_{2}}$ is the remaining length of the connection after SU resumes connection at time $t_2$. Since either a newly generated SU connection after resumption, or the remaining part of a preempted connection needs a new connection arrangement on the control channel, the number of slots occupied by each connection type is the same.
\end{proof}
\label{prop:memory}
\end{observation}

Having Observation~\ref{prop:memory} we can derive transition probabilities. Because packet generation is affected by the number of connections, we use $Z_t$ to classify conditions to derive the state transition probabilities. Due to the assumption of a maximum number of one connection generation in one time slot, the state transition probability of the case of $z>m+1$ is zero.

For $z \le m+1$, a data connection is terminated only if the transmitting node completes its transmission without PU interruption. When a PU is detected in the channel, the SU temporarily pauses communication without terminating the connection. Among $z$ SU connections, $x$ connections actually utilize channels for data transmission and $z-x$ connections are paused due to PU detection. Thus $\binom{z}{z-x}$ combinations of PU appearance are possible. At the same time, because the total number of channels occupied by PUs is $y$, the remaining $y-z+x$ PUs should appear on $M_D-z$ idle channels. Thus there can be $\binom{M_D-z}{y-z+x}$ combinations for PU appearance on idle channels.

The SU connection generation probability for $z=m+1$ is $T_{k}^{(0)}S_{m}^{(1)}$, just like in the case B$_0$S$_0$. For $z<m+1$ (i) no connection is generated and $m-z$ connections are terminated, and (ii) one connection is generated and $m-z+1$ connections are terminated. For $z=m+1$, one connection is generated while no connections are terminated. 

For the transition between full connection states, i.e. the state transition from $m=s_m$ to $z=s_m$, we should take into account the case that one connection is generated and no connections are terminated because there is no available resources for a new connection.

Finally, considering all these cases and defining $R_{x,y}^{(z)}=\binom{z}{z-x}\binom{M_D-z}{y-z+x} p_c^y(1-p_c)^{M_D-y}$ the state transition probability is given as
\begin{equation}
p_{xyz|klm}=
\begin{cases}
0, & z>m+1,\\
T_{k}^{(0)}S_{m}^{(1)}R_{x,y}^{(z)}, & z=m+1,\\
\left(T_k^{(m-z)} S_m^{(0)} + T_k^{(m-z+1)} S_m^{(1)}\right)R_{x,y}^{(z)}, & z < m+1, m<s_m\text{~or~}z<s_m,\\
\left(T_k^{(0)} S_m^{(0)} + T_k^{(1)} S_m^{(1)} + T_k^{(0)} S_m^{(1)}\right)R_{x,y}^{(z)}, & z=m=s_m.\\
\end{cases}
\label{eq:p_buf_noswc}
\end{equation}
Note that this OSA MAC has been previously analyzed in~\cite{pawelczak_tvt_2009}. As it has been pointed out, the model proposed did not work well for the full range of parameters. This is due to the following. A Markov model has been derived for $\{X_t,Y_t\}$ (using unmodified transition probabilities of~\cite[Eq.~6]{mo_tmc_2008} used to calculate average throughput of networks based on non-OSA multichannel MAC protocols). With this limitation termination, the probability in~\cite[Eq. (14)]{pawelczak_tvt_2009}, analogue to (\ref{eq:Tkj}), included an aggregated stream of PU and SU traffic, where PU traffic $q_p$ was later substracted from steady state channel utilization in~\cite[Eq. (10)]{pawelczak_tvt_2009}, analogue to (\ref{eq:R_micro}). The approximation of~\cite{pawelczak_tvt_2009}, although Markovian, worked reasonably well only for a moderate values of PU activity $q_p$\footnote{November 29, 2012: After another verification of the analytical model of Pawelczak et al.~\cite{pawelczak_tvt_2009} with that of ours it appears that the model for B$_1$S$_0$ MAC of this paper actually matches the one of~\cite{pawelczak_tvt_2009} (assuming no capture effect and $M_D=M-1$ for~\cite{pawelczak_tvt_2009}). Therefore this observation is therefore incorrect. Matlab code that verifies the above statement (implementing both analytical models) is available upon request.}.

\subsubsection{Case B$_1$S$_1$}
\label{sec:BS}

This OSA MAC from analysis perspective is the same as the buffering OSA MAC with no channel switching, except for the following two differences. First, if there is at least one idle channel, an SU that has a connection but does not utilize a channel cannot exist because this SU can switch to the idle channel. Formally, the state transition to the state of $z\ne x$ and $x+y<M_D$ and the state transition from the state $m\ne k$ and $k+l<M_D$ are not possible. Second, in contrast to the non-switching OSA MAC, $y$ PUs can appear in any of $M_D$ channels because the SUs can switch to the idle channels in this option, the same as for B$_0$S$_1$ case. Thus the possible number of cases of PU appearance is just $\binom{M_D}{y}$. Therefore, replacing $R_{x,y}^{(z)}$ with $R_{0,y}^{(0)}$ and adding conditions $z\ne x, x+y<M_D, m\ne k$ and $k+l<M_D$ to the condition $z>m+1$ in (\ref{eq:p_buf_noswc}) results in a complete definition of $p_{xyz|klm}$. For consistency we present this state transition probability as
\begin{equation}
p_{xyz|klm}=
\begin{cases}
0, &\begin{split}&z>m+1, \textrm{or~} z\ne x, x+y<M_D,\\ &\qquad \textrm{or~} m\ne k, k+l<M_D,\end{split}\\
T_{k}^{(0)}S_{m}^{(1)}R_{0,y}^{(0)}, & z=m+1,\\
\left(T_k^{(m-z)} S_m^{(0)} + T_k^{(m-z+1)} S_m^{(1)}\right)R_{0,y}^{(0)}, & z < m+1, m<s_m \textrm{~or~}z<s_m,\\
\left(T_k^{(0)} S_m^{(0)} + T_k^{(1)} S_m^{(1)} + T_k^{(0)} S_m^{(1)}\right)R_{0,y}^{(0)}, & z=m=s_m.\\
\end{cases}
\label{eq:p_buf_swc}
\end{equation}

\subsubsection{Impact of Channel Error on the Throughput Calculations}
\label{sec:Error}

All previous analysis were done under the assumption of the error-free channel. In this section we will  briefly discuss the impact of channel error on the throughput calculations. 

Channel error impacts the throughput in two ways. First, error affects throughput when SU involved in a connection setup fails to receive a control message from the transmitter. As a result no connection is established. Second, error affects throughput when SU not associated with the current connection setup (which does not overhear the connection setup from others) collide later with other users, believing incorrectly it selected a free data channel for communication. The throughput of DCC is impacted by both effects. On the other hand, HCC is influenced only by the first effect. It is because HCC MAC protocol implementations do not posses a separate control channel. Thus no overhearing of connection setup is possible and users with HCC MAC protocol select data channel for communication automatically according to a pre-defined hopping sequence. Because of the prohibitive complexity of the analysis of the second effect, we focus on the first error case and the HCC.

For HCC, the control channel is selected as one of the data channels by a hopping method. Thus, if we assume an error on the control channel, it is reasonable to consider the error on the data channel as well. For the control channel, if an error occurs, a connection fails to be established. Thus it is modeled by multiplying $\hat{S}_{m}$ by $1-p_e$, where $p_e$ is a probability of error in the current time slot. For the data channel, different error handling strategies can be considered. We focus on the two following situations: i) case E$_1$ denoting packet punctured by unrecovered errors and ii) case E$_2$ denoting transmission termination on error. 

\paragraph{Case  E$_1$}

It can be assumed that when an error occurs on a time slot, the SU simply discards that time slot and resumes transmitting the remaining packet fragment from the next correct time slot. This is modeled by replacing the capacity $C$ with $C(1-p_e)$. 

\paragraph{Case  E$_2$}

It can also be assumed that the connection terminates when an error occurs. Thus the probability that the packet finishes transmitting, $q$, should be replaced by $q+(1-q)p_e$. In addition, if the control channel hops to a channel which is being utilized for data transmission but error occurs, a new connection cannot be established. This is modeled by multiplying $\hat{S}_{m}$ by $(1-p_e)^2$.

\section{Numerical Results}
\label{sec:numerical_results}

We now present numerical results for our model. First, we present results independently for spectrum sensing and OSA MAC performance, in Section~\ref{sec:spectrum_sensing_performance} and Section~\ref{sec:mac_performance}, respectively, for the microscopic case. Then in Section~\ref{sec:joint_performance} we present the results of joint optimization of these two layers in the microscopic and macroscopic context. Moreover, due to a vast combination of parameters to consider we have decided to follow the convention of~\cite{pawelczak_tvt_2009,mo_tmc_2008} and focus on two general network setups (unless otherwise stated): (i) small scale network with $M=3$, $N=12$, $d=5$\,kB and (ii) large scale network with $M=12$, $N=40$, $d=20$\,kB.

In this section we will also compare the analytical model of the sensing layer and OSA MAC protocols to simulation results. The simulations were developed with Matlab and reflect exactly the sensing models and MAC protocols presented in this paper. Simulation results for each system were obtained using the method of batch means for a 90\% confidence interval. To evaluate the sensing protocols each batch contained 100 events and the whole simulation run was divided into 10 batches with no warm up phase. When simulating the OSA MAC protocols, each batch contained 1000 events while the whole simulation was divided into 100 batches with the warm up period equal of 100 events.

\subsection{Spectrum Sensing Architecture Performance}
\label{sec:spectrum_sensing_performance}

For all possible combinations of sensing architectures we compute the probability of false alarm for a wide range of $t_q$. For two networks considered we select a set of the following common parameters: $t_t=t_{d,\max}=1$\,ms, $C=1$\,Mbps, $b=1$\,MHz, $q_p=0.1$ (which approximately corresponds to the level of actual measured PU occupancy on the channel from~\cite[Tab. 1]{pawelczak_wcm_2009}), $\gamma=-5$\,dB, and $\alpha=1/M$. In all sections, except for Section~\ref{sec:impact_sensing_error}, we present results for an error-free channel. Note that for all results presented in this section simulation results for all protocols confirm the accuracy of the analytical model.

\subsubsection{Measurement Reporting Protocol Performance}
\label{sec:measurement_reporting}

\begin{figure}
\centering
\subfigure[]{\includegraphics[width=0.49\columnwidth]{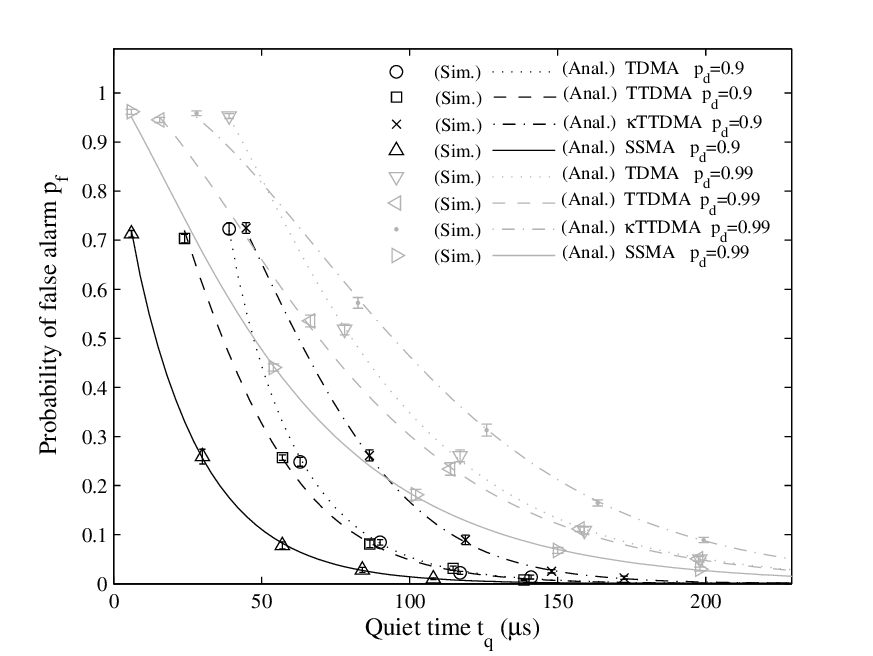}\label{fig:M3N12pdOH}}
\subfigure[]{\includegraphics[width=0.49\columnwidth]{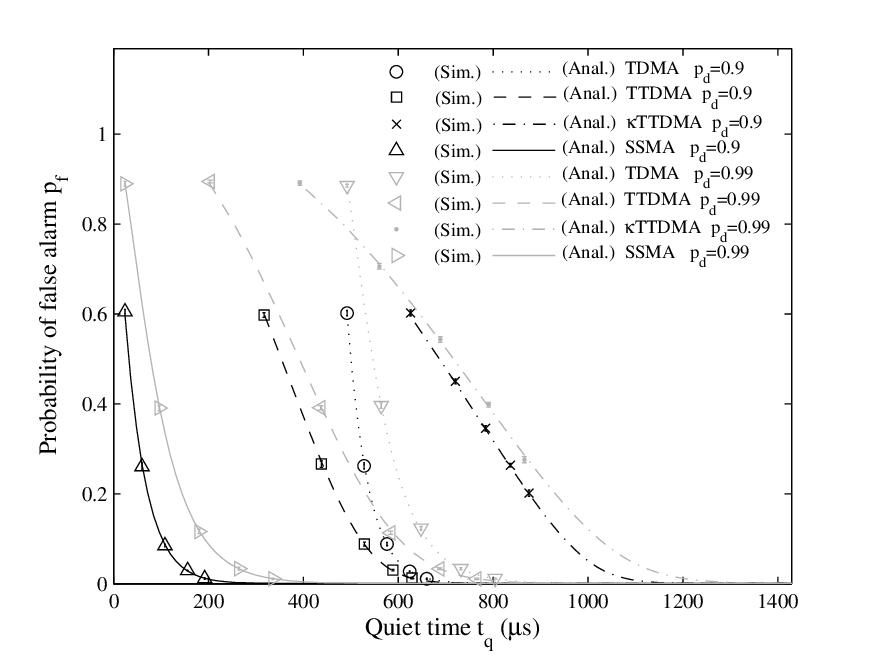}\label{fig:M12N40pdOH}}
\caption{Performance of different measurement reporting protocols as a function of $t_q$ and $p_d=p_{d,\min}$ for (a) $M=3$, $N=12$, and (b) $M=12$, $N=40$. Common parameters: $p_e=0$, $\kappa=1$ (``or'' rule), $t_t=t_{d,\max}=1$\,ms, $C=1$\,Mbps, $b=1$\,MHz, $q_p=0.1$, $\gamma=-5$\,dB, $\alpha=1/M$, and $n_g=1$.}
\label{fig:sensing_gr1}
\end{figure}

The results are presented in Fig.~\ref{fig:sensing_gr1}. For the same $p_d$ requirement, SSMA results in the lowest $p_f$ for each value of $t_q$, while TDMA performs worst. The benefit of introducing TTDMA in comparison to TDMA is clearly visible for all network scenarios. 

The advantage of TTDMA and SSMA can be shown more clearly if we compare the results of different $p_d=p_{d,\min}$ requirements. We can observe that high detection requirement such as $p_d=0.99$ makes the performance worse, as generally known. However if TTDMA or SSMA is applied, the performance for $p_d=0.99$ can be higher than that of TDMA for $p_d=0.9$. For example, in the range that $t_q<50$\,$\mu$s in Fig.~\ref{fig:M3N12pdOH}, SSMA for $p_d=0.99$ outperforms TDMA for $p_d=0.9$. Moreover, in Fig.~\ref{fig:M12N40pdOH}, for $t_q\lessapprox550$\,$\mu$s, SSMA and TTDMA for $p_d=0.99$ outperforms TDMA for $p_d=0.9$. 

It is important to note that $\kappa$TTDMA performs worse than the rest of the protocols. It is due to excessive delay caused by instant acknowledgment of reporting result to the cluster head node. Note that $\kappa$TTDMA is a lower bound for the operation of TTDMA. Also note that when TDMA needs to be equipped with acknowledgment function, as $\kappa$TTDMA, its performance would be degraded the same way as TTDMA. Since we analyze static network with pre-set parameter values, e.g. $\kappa$ does not change over time, in the following sections we proceed with unmodified TTDMA only.

\subsubsection{Impact of Channel Errors during Reporting on PU Detection Performance}
\label{sec:impact_sensing_error}

\begin{figure}
\centering
\subfigure[]{\includegraphics[width=0.49\columnwidth]{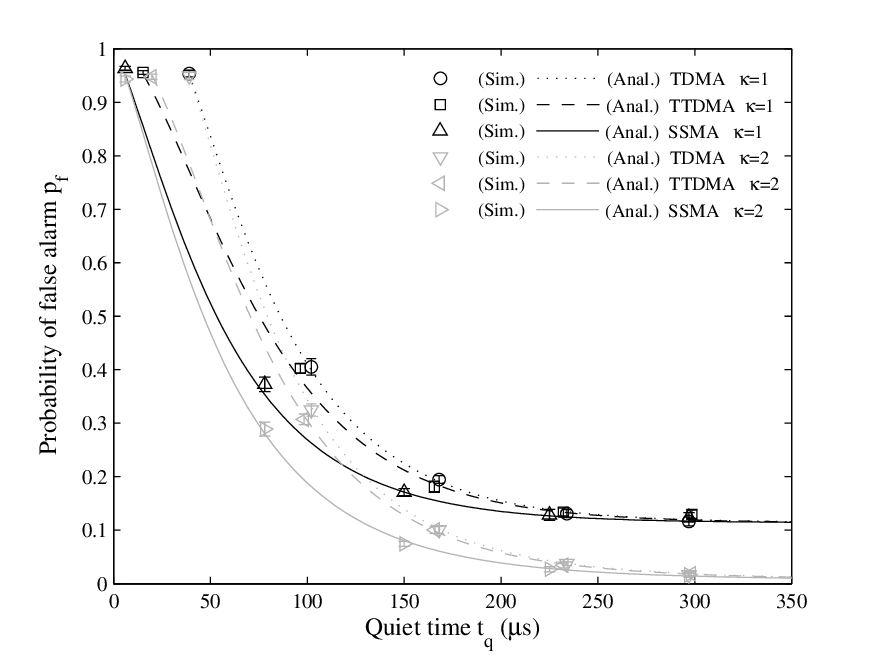}\label{fig:M3N12pd_error}}
\subfigure[]{\includegraphics[width=0.49\columnwidth]{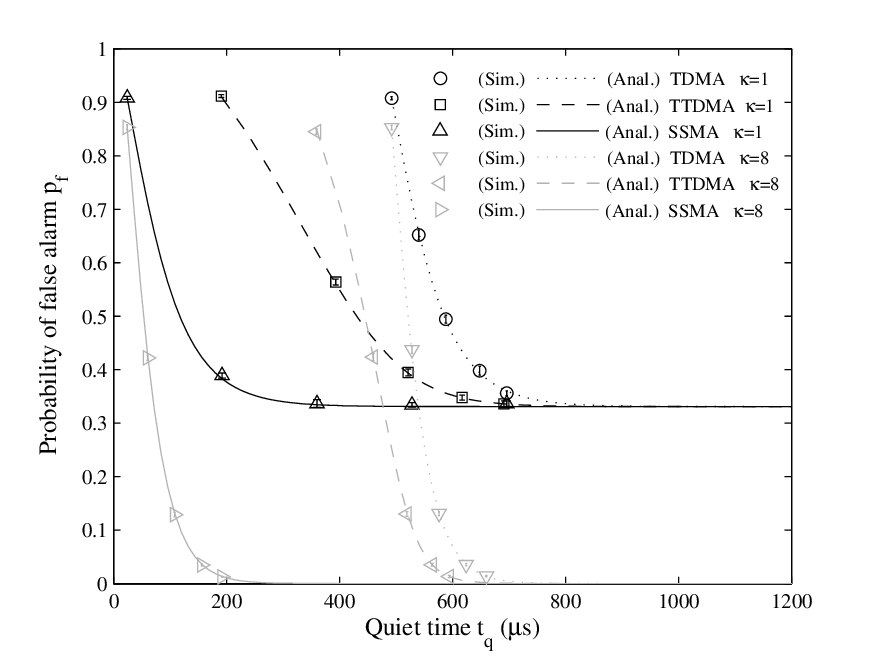}\label{fig:M12N40pd_error}}
\caption{The effect of channel errors on the performance of different measurement reporting protocols as a function of $t_q$ and $p_d=p_{d,\min}=0.99$ for (a) $M=3$, $N=12$, $\kappa=\{1,2\}$, and (b) $M=12$, $N=40$, $\kappa=\{1,8\}$. All remaining parameters are the same as in Fig.~\ref{fig:sensing_gr1} except for $p_e=0.01$ .}
\label{fig:sensing_error}
\end{figure}

The results are presented in Fig.~\ref{fig:sensing_error}. For small and large scale network, and the same parameters as used in Section~\ref{sec:measurement_reporting}, we have observed the probability of false alarm keeping detection probability $p_d$ constant for varying quiet time $t_q$. 
First, it is obvious when comparing Fig.~\ref{fig:sensing_gr1} (no channel error) and Fig.~\ref{fig:sensing_error} (channel error) the impact of error is clearly visible, i.e. $p_f$ increases for every protocol. However, the relation between individual protocols is the same since error affects all protocols equally. Second, the effect of error on the small scale network is smaller than for the large scale network, compare Fig.~\ref{fig:M3N12pd_error} and Fig.~\ref{fig:M12N40pd_error}, since the probability that SU will send a wrong report is larger for network with large number of nodes. Lastly, for small values of $\kappa$ probability of false alarm stabilizes and never reaches zero. However, large values of $\kappa$ reduce significantly the effect of channel errors. It is because with high $\kappa$ probability of making an error decreases rapidly. With 20\% of nodes participating in the cooperative agreement on PU state, $\kappa=2$ for small network and $\kappa=8$ for large scale network, effect of error is reduced almost to zero.

\subsubsection{Impact of Cooperation Level on PU Detection Performance}
\label{sec:kappa}

\begin{figure}
\centering
\subfigure[]{\includegraphics[width=0.49\columnwidth]{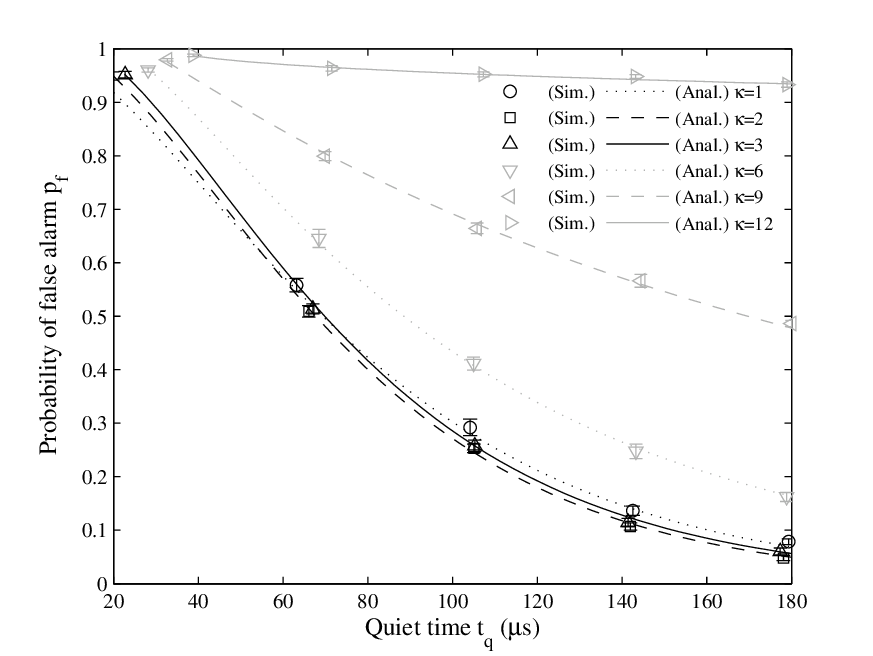}\label{fig:M3N12k}}
\subfigure[]{\includegraphics[width=0.49\columnwidth]{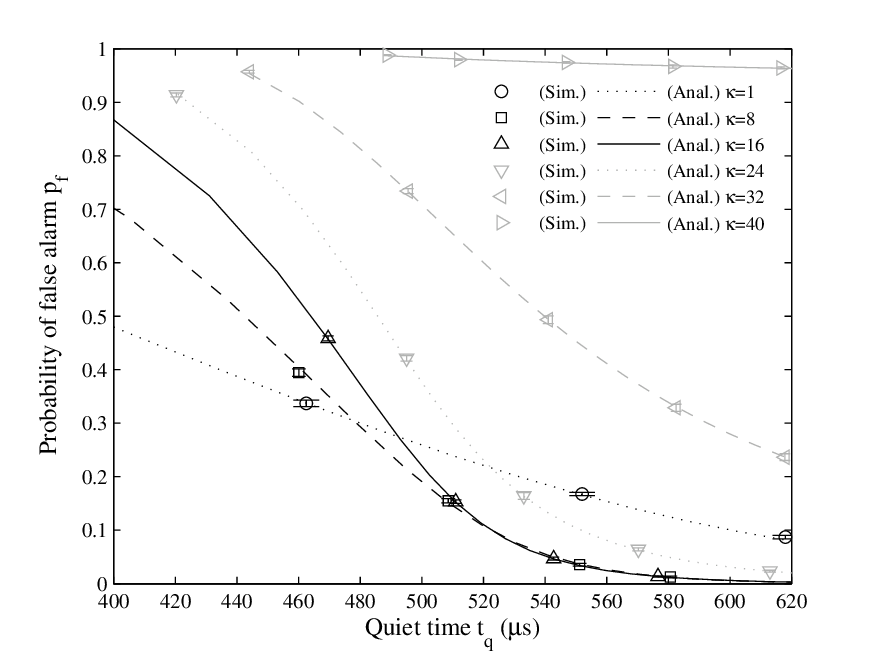}\label{fig:M12N40k}}
\caption{Performance of TTDMA as a function of $t_q$ and $\kappa$ for (a) $M=3$, $N=12$, and (b) $M=12$, $N=40$. Rest of the parameters are the same as in Fig.~\ref{fig:sensing_gr1}, except for $p_{d,\min}=0.99$.}
\label{fig:sensing_gr2}
\end{figure}

The results are presented in Fig.~\ref{fig:sensing_gr2}. We have selected TTDMA and set $p_d=p_{d,\min}=0.99$ as a protocol for further investigation. We observe that for the small scale network, see Fig.~\ref{fig:M3N12k}, the performance for $\kappa=2$ is the best, while for the large scale network, see Fig.~\ref{fig:M12N40k}, the best performance can be achieved when $\kappa=8$ or $16$ if $p_f<0.1$. Based on this observation, we conclude that for given detection requirements, high detection rate of PU is obtained when $\kappa$ is well below the total number of SUs in the network. While for the considered setup optimal $\kappa\approx20\%$, this value might be different for other network configurations.

\subsubsection{Impact of Group Size on PU Detection Performance}
\label{sec:group_selection}

\begin{figure}
\centering
\subfigure[]{\includegraphics[width=0.49\columnwidth]{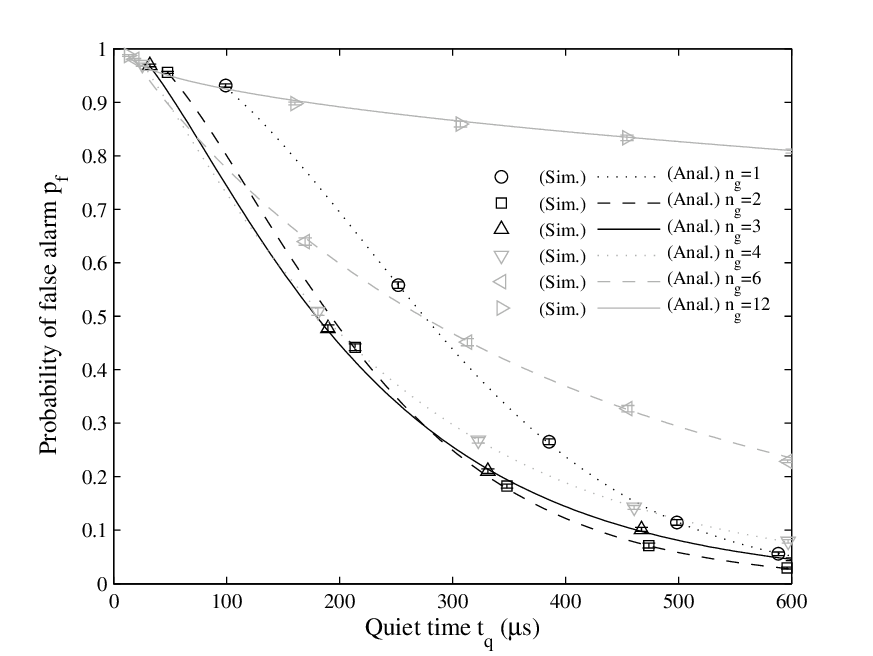}\label{fig:M12N20ng}}
\subfigure[]{\includegraphics[width=0.49\columnwidth]{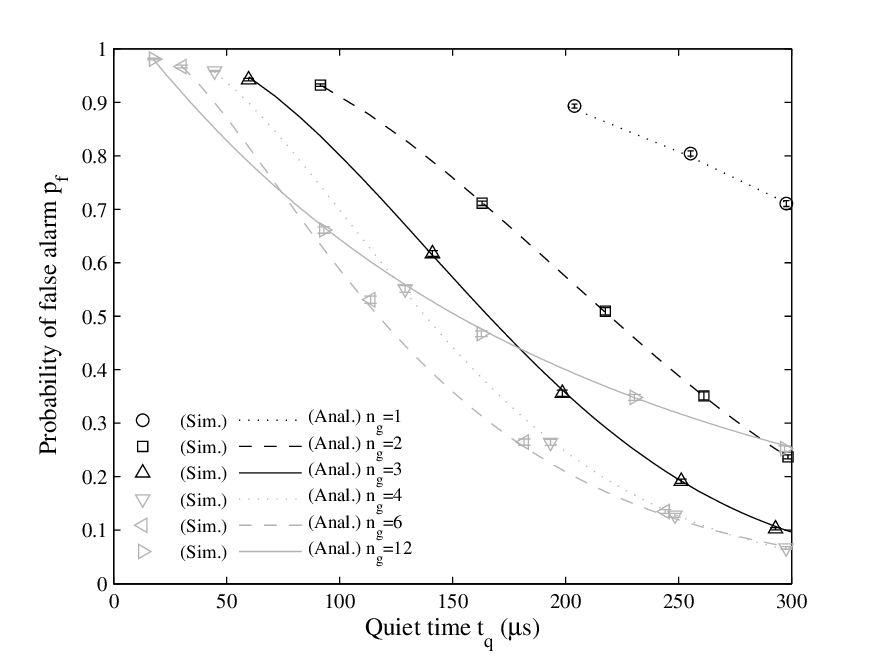}\label{fig:M12N40ng}}
\caption{Performance of TTDMA as a function of $t_q$ and $n_g$ for (a) $M=12$, $N=20$, and (b) $M=12$, $N=40$. Rest of the parameters are the same as in Fig.~\ref{fig:sensing_gr1}, except for $p_{d,\min}=0.99$.}
\label{fig:sensing_gr3}
\end{figure}

The results are presented in Fig.~\ref{fig:sensing_gr3}. To contrast the impact of group size, we choose $M=12$, $N=20$ as the small scale network. We perform experiments only for the case when $\bar{m}_s$ equal for all groups, which means the number of groups is the divisor of $M$, i.e. $n_g \in \{1,2,3,4,6,12\}$. 

An interesting observation is that the number of groups to achieve the best performance becomes larger as the number of users $N$ increases. For the small scale network, see Fig.~\ref{fig:M12N20ng}, the best performance is observed for $n_g=2$ or $n_g=3$, while for large scale network, Fig.~\ref{fig:M12N40ng}, $n_g=6$ is the best. This is because for the large scale network, the reporting overhead caused by large number of users offsets the performance improvement achieved by large cooperation scale.

\subsubsection{Impact of $\kappa$ on PU Detection Performance}
\label{sec:imact_kappa}

\begin{figure}
\centering
\subfigure[]{\includegraphics[width=0.49\columnwidth]{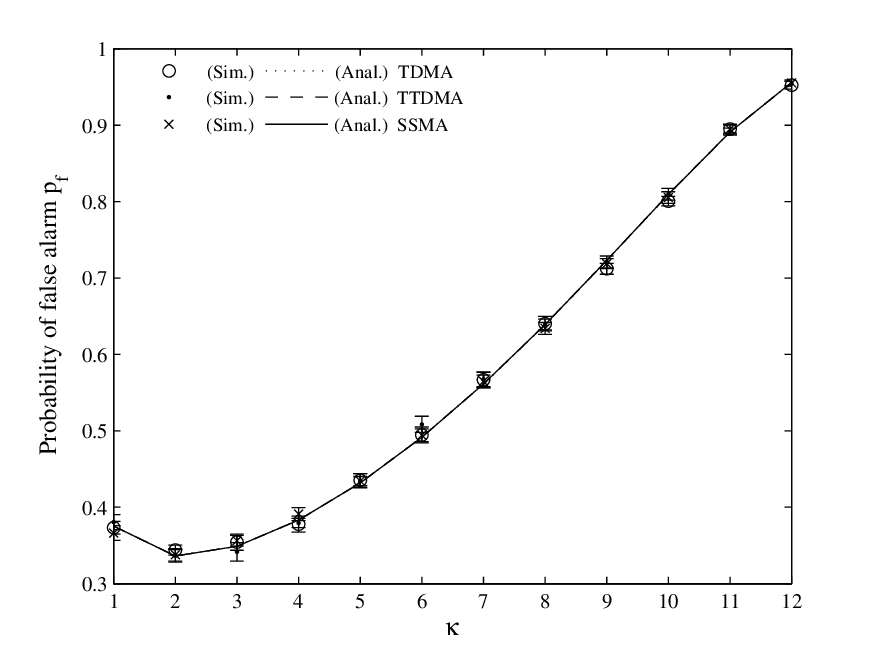}\label{fig:M3N12k_pf}}
\subfigure[]{\includegraphics[width=0.49\columnwidth]{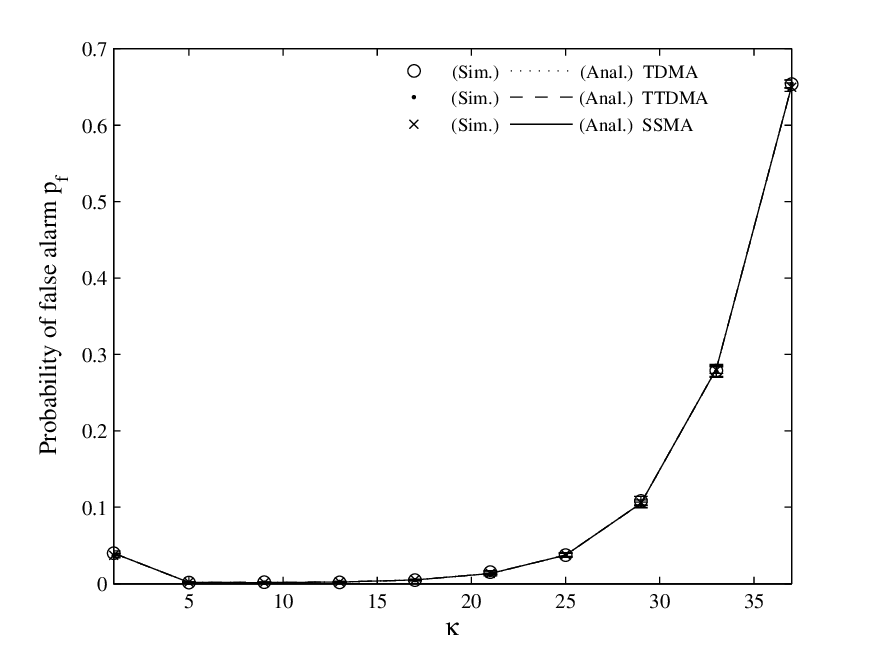}\label{fig:M12N40k_pf}}
\subfigure[]{\includegraphics[width=0.49\columnwidth]{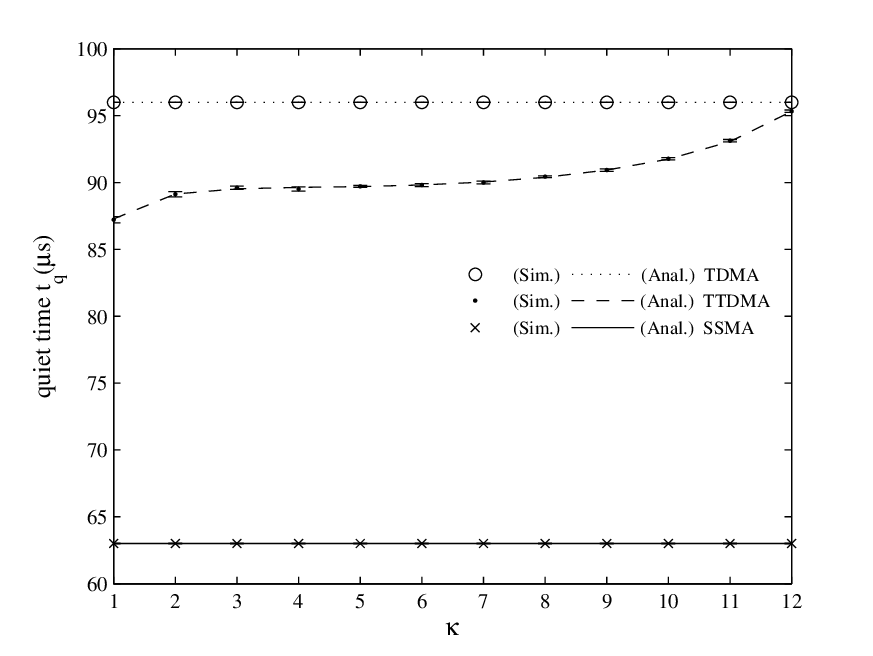}\label{fig:M3N12k_tq}}
\subfigure[]{\includegraphics[width=0.49\columnwidth]{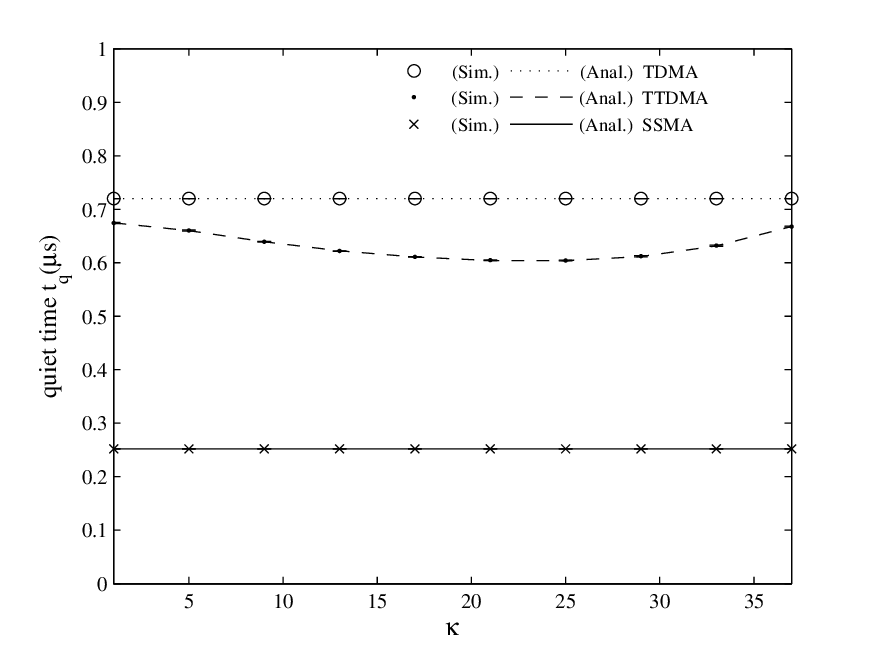}\label{fig:M12N40k_tq}}
\caption{The effect of varying $\kappa$ on (a), (b) false alarm probability, and (c), (d) quiet time of different measurement reporting protocols as a function for $p_d=p_{d,\min}$ and (a), (c) small scale network, $M=3$, $N=12$, and (b), (d) large scale network, $M=12$, $N=40$. Rest of the parameters are the same as in Fig.~\ref{fig:sensing_gr1} except for $t_e=20$\,ms.}
\label{fig:sensing_kappa}
\end{figure}

The results are presented in Fig.~\ref{fig:sensing_kappa}. For two network sizes, large and small, three sensing MAC protocols and fixed value of $p_d$ we vary $\kappa$ to see its impact on the sensing MAC protocol performance. We observe probability of false alarm $p_f$ as well as change in quite time $t_q$. First we notice that varying $\kappa$ does not change probability of false alarm for any protocol, in both network configurations. Moreover the lowest probability of false alarm is obtained when small number of users agree on the PU state. The larger the channel number, the larger the range of $\kappa$ when network obtains the lowest probability of false alarm, compare Fig.~\ref{fig:M3N12k_pf} and Fig.~\ref{fig:M12N40k_pf}. The trends of $p_f$ for both network configurations are the same, since all protocols keep false alarm rate on the same level irrespective of the parameter change. In case of quiet time, TDMA and SSMA have $q_p$ constant and independent  from $\kappa$, which differs them from TTDMA whose operation strictly depends on the value of $\kappa$ considered. And again, when comparing Fig.~\ref{fig:M3N12k_tq} and Fig.~\ref{fig:M12N40k_tq} the optimal value of $t_q$ for TTDMA is in the same range as $p_f$ which proves the optimality of the design.

\subsection{OSA MAC Protocol Performance}
\label{sec:mac_performance}

To evaluate the effectiveness of all proposed and analyzed MAC protocols we have fixed $C=1$\,Mbps, $p=e^{-1}/N$, $t_q=t_p=100$\,$\mu$s, $t_t=t_{d,\min}=1$\,ms, $p_{d,\min}=p_d=0.99$, and $p_f=0.1$. Note that we do not relate $p_d$ and $p_f$ with the actual spectrum sensing process at this moment (this will be done in Section~\ref{sec:joint_performance}), assuming that spectrum sensing layer is able to obtain such quality of detection. Again, as in Section~\ref{sec:spectrum_sensing_performance}, results are presented separately for error-free and error channel. 

\subsubsection{Impact of PU Activity Level on OSA MAC Protocols}
\label{sec:pu_level_impact}

\begin{figure}
\centering
\subfigure[]{\includegraphics[width=0.49\columnwidth]{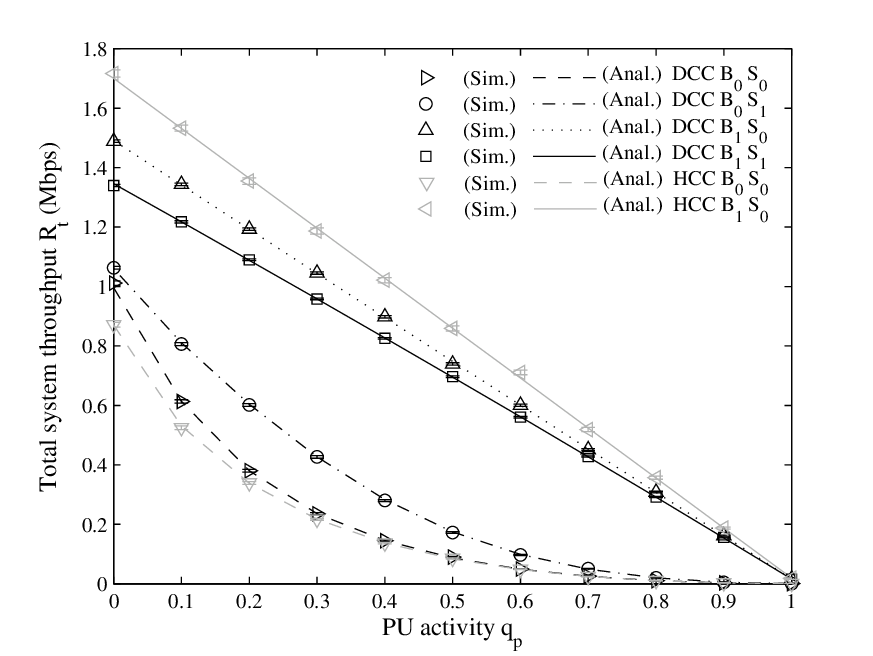}\label{fig:}}
\subfigure[]{\includegraphics[width=0.49\columnwidth]{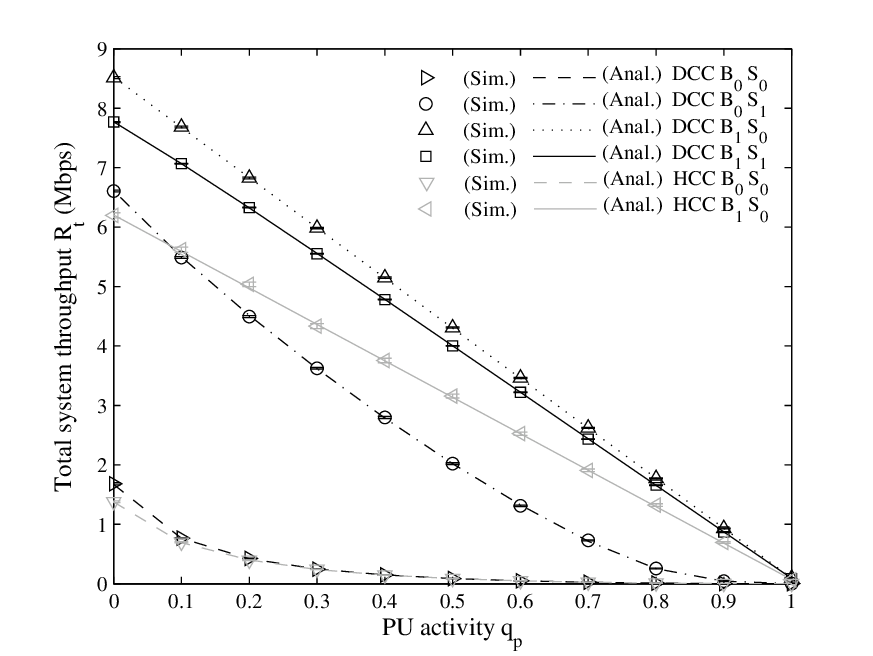}\label{fig:}}
\caption{Performance of OSA MAC protocols versus PU activity level for (a) $M=3$, $N=12$, $d=5$\,kB and (b) $M=12$, $N=40$, $d=20$\,kB. Common parameters: $p_e=0$, $p=e^{-1}/N$, $p_d=p_{d,\min}=0.99$, $p_f=0.1$, $t_q=t_p=100$\,$\mu$s, $t_t=t_{d,\max}=1$\,ms, and $C=1$\,Mbps.}
\label{fig:q_p_impact}
\end{figure}

The results are presented in Fig.~\ref{fig:q_p_impact}. We observe that PU activity degrades DCC and HCC for B$_{0}$S$_{0}$, irrespective of other network parameters. Their performances are comparable in this case. DCC and HCC performs best with B$_{1}$S$_{0}$.  The results show that the non-buffering OSA MAC protocols are very sensitive to $q_p$ where the greatest throughput decrease is visible at low ranges of PU activity. On the other hand, with connection buffering we observe a linear relation between $q_p$ and $R_t$.

\subsubsection{Impact of SU Packet Size on OSA MAC Protocols}
\label{sec:su_packet_impact}

\begin{figure}
\subfigure[]{\includegraphics[width=0.49\columnwidth]{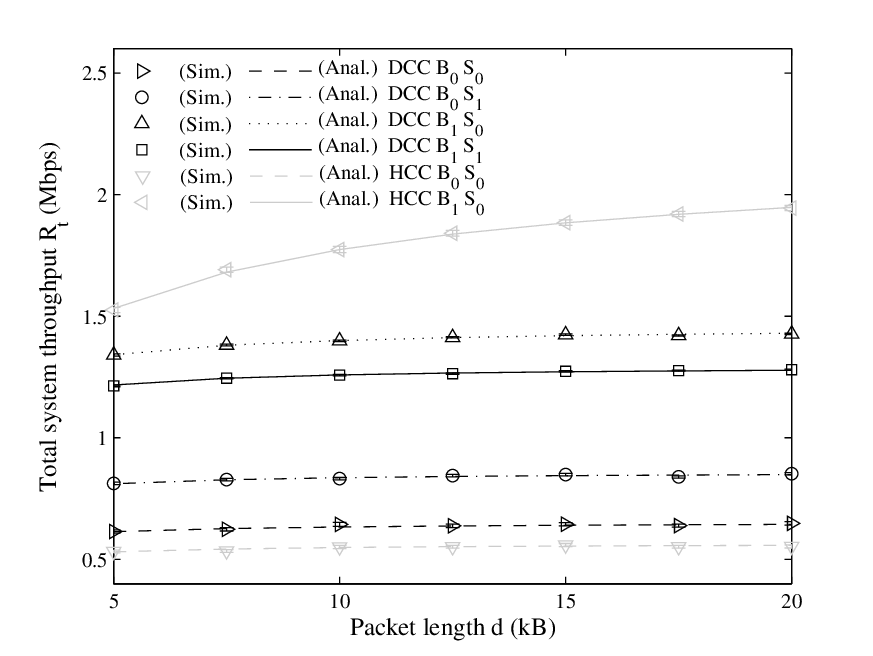}\label{fig:su_packet_impact1}}
\subfigure[]{\includegraphics[width=0.49\columnwidth]{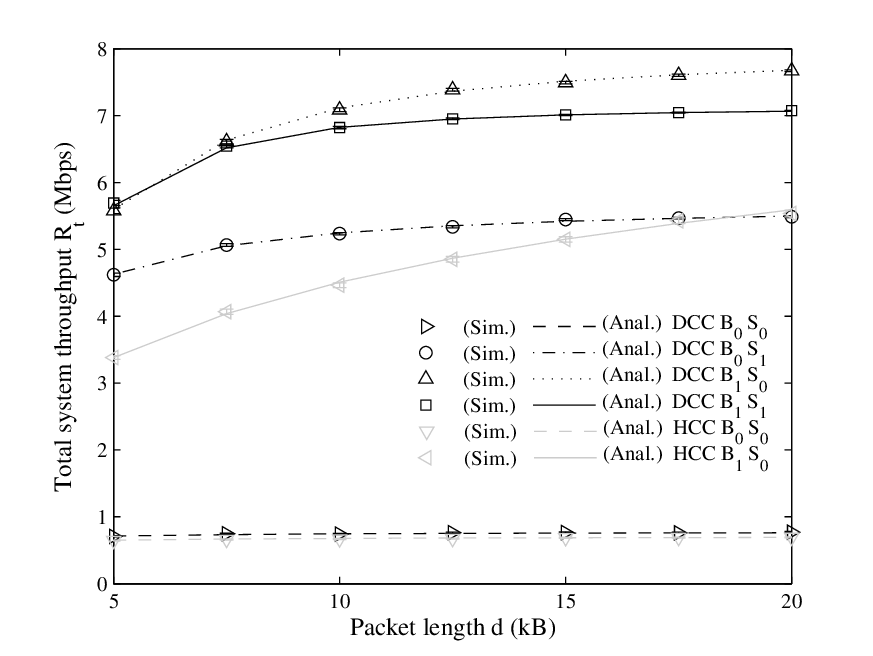}\label{fig:su_packet_impact2}}
\caption{Performance of OSA MAC protocols versus packet size $d$ for (a) $M=3$, $N=12$, and (b) $M=12$, $N=40$. Rest of the parameters are the same as in Fig.~\ref{fig:q_p_impact}, except for $q_p=0.1$.}
\label{fig:su_packet_impact}
\end{figure}

The results are presented in Fig.~\ref{fig:su_packet_impact}. Obviously, for larger SU packet size, the OSA network is able to grab more capacity. However, when packets become excessively large the throughput saturates. It remains that with no buffering and no channel switching protocols obtain the lowest throughput, no matter what network setup is chosen. Interestingly, although intuitevely B$_{1}$S$_{1}$ should obtain the highest channel utilization, it does not perform better than B$_{1}$S$_{0}$ due to large switching time. With $t_p$ approaching zero, DCC B$_{1}$S$_{1}$ would perform best, irrespective of the network setup as we discuss below.

\subsubsection{Impact of Switching Time on OSA MAC Protocols}
\label{sec:switching_impact}

\begin{figure}
\subfigure[]{\includegraphics[width=0.49\columnwidth]{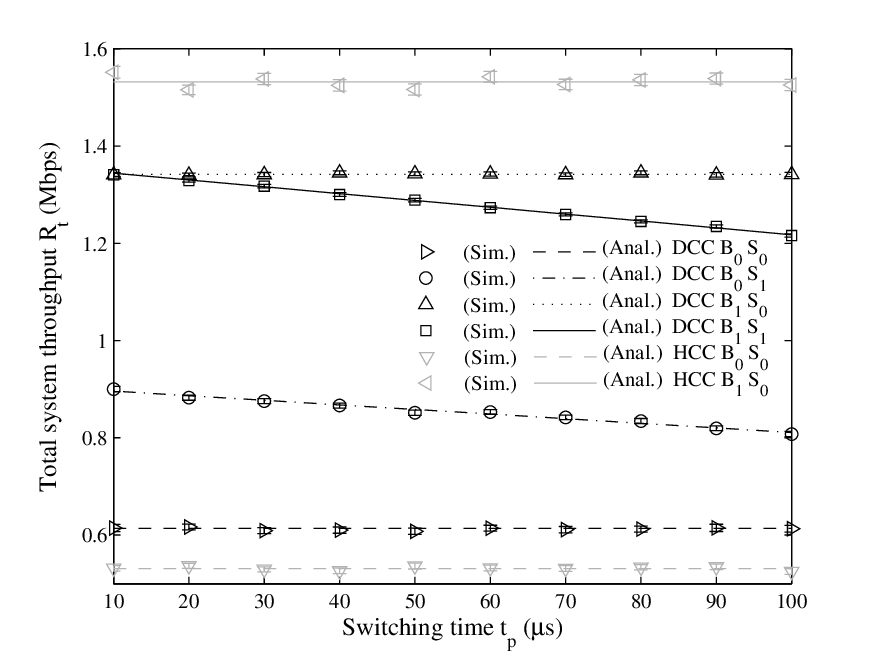}\label{fig:tp_M3N12}}
\subfigure[]{\includegraphics[width=0.49\columnwidth]{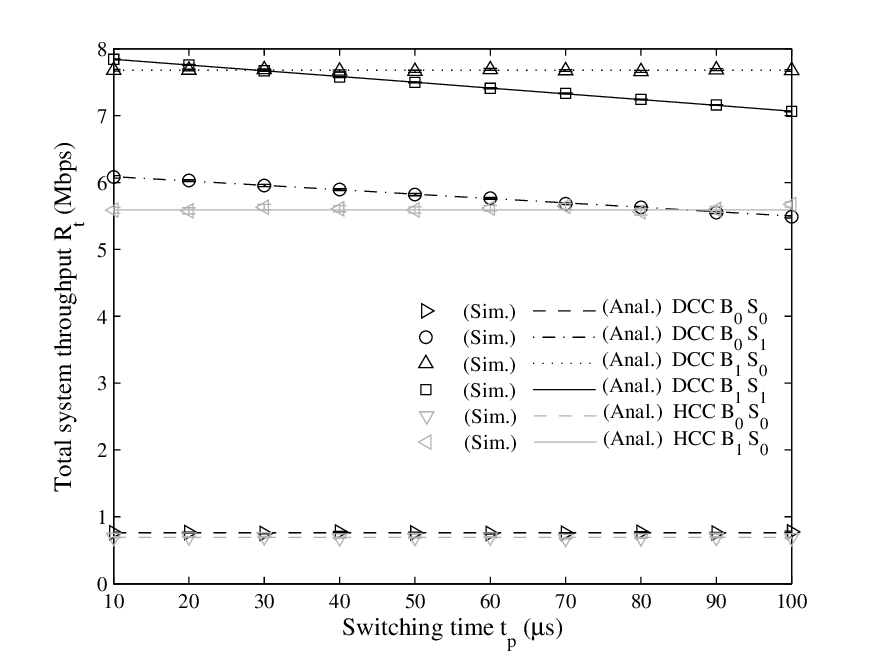}\label{fig:tp_M12N40}}
\caption{Performance of OSA MAC protocols versus channel switching time $t_p$ for (a) $M=3$, $N=12$, $d=5$\,kB and (b) $M=12$, $N=40$, $d=20$\,kB. Rest of the parameters are the same as in Fig.~\ref{fig:q_p_impact}, except for $q_p=0.1$.}
\label{fig:t_p_impact}
\end{figure}
The results are presented in Fig.~\ref{fig:t_p_impact}. In this experiment, we verify that for small $t_p$ DCC B$_{1}$S$_{1}$ outperforms DCC B$_{1}$S$_{0}$. However, there is no huge difference between their performances even at $t_p=10$\,$\mu$s. This is because connection switching does not seriously impact the data throughput for the network setups in which the number of channels is less than the number of possible connections, i.e. $M<2N$. For this network setup, all channels are utilized in most of time, and therefore there may not exist many idle channels to switch. The performance of DCC B$_{0}$S$_{1}$ is also improved for small $t_p$, and we observe that DCC B$_{0}$S$_{1}$ outperforms HCC B$_{1}$S$_{0}$ for large scale network, see Fig.~\ref{fig:tp_M12N40}.

\subsubsection{Relation Between Number of SUs and PU Channels}
\label{sec:su_packet_impact}

\begin{figure}
\subfigure[]{\includegraphics[width=0.49\columnwidth]{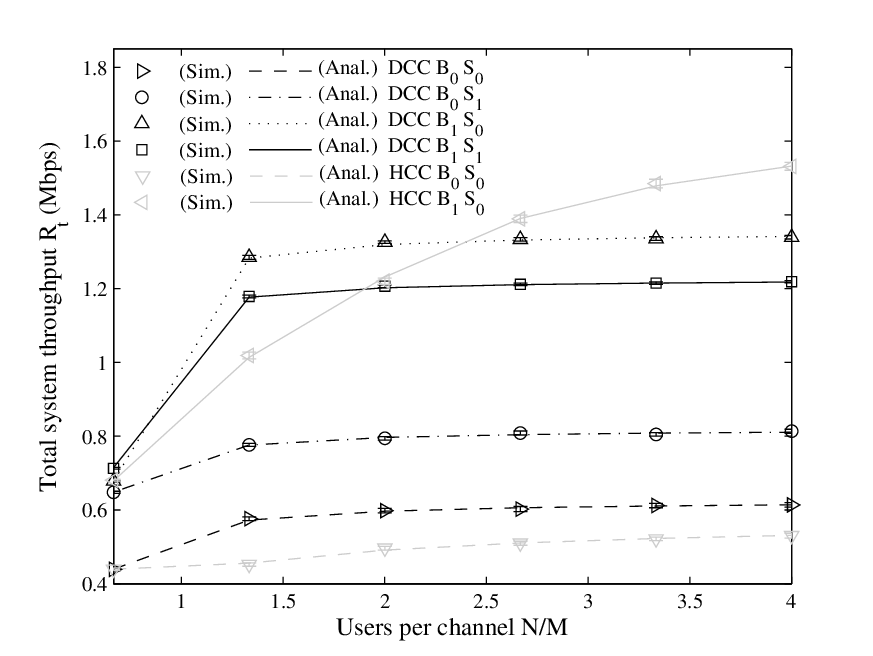}\label{fig:NM_M3qp01pl5}}
\subfigure[]{\includegraphics[width=0.49\columnwidth]{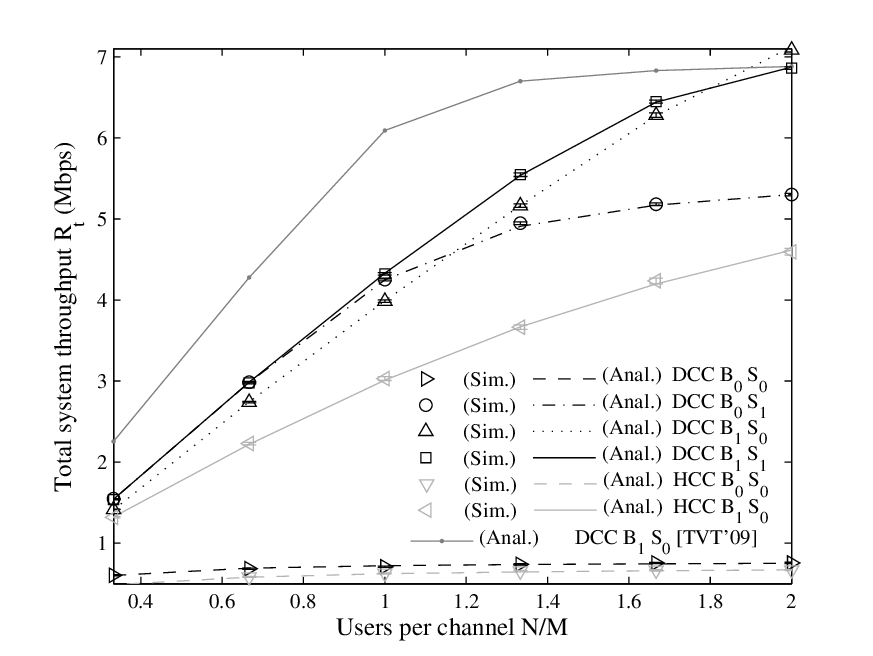}\label{fig:NM_M12qp01pl20}}
\caption{Performance of OSA MAC protocols for ratio of number of SUs to number of PU channels for (a) $M=3$, $d=5$\,kB and (b) $M=12$, $d=20$\,kB. Rest of the parameters are the same as in Fig.~\ref{fig:q_p_impact}, except for $q_p=0.1$.}
\label{fig:M_N_impact}
\end{figure}

Finally we want to explore the relationship between the number of OSA network users and the number of available PU channels. The results are presented in Fig.~\ref{fig:M_N_impact}.

With increasing ratio $N/M$ we observe an increasing throughput, where at some point all protocols almost saturate. Again because of the high switching penalty DCC with  B$_{1}$S$_{1}$ is inferior to B$_{1}$S$_{0}$. For small scale network, as shown in Fig.~\ref{fig:NM_M3qp01pl5}, a separate comment is needed for HCC B$_{1}$S$_{0}$. For small $N/M$ DCC with B$_{1}$S$_{1}$ and B$_{1}$S$_{0}$ obtains higher throughput than HCC B$_{1}$S$_{0}$. However, for high $N/M$ HCC B$_{1}$S$_{0}$ achieves the highest $R_t$ of all protocols. For large scale network as shown in Fig.~\ref{fig:NM_M12qp01pl20}, comparing channel switching and buffering options we conclude that much more channel utilization is obtained by connection buffering than by channel switching alone when $N/M>1$.

Note that for all cases described in this section simulation results agrees with our analytical model. Comparing our model and analytical results of~\cite{pawelczak_tvt_2009} for DCC B$_1$S$_0$, see Fig.~\ref{fig:NM_M12qp01pl20}, we observe that prior analysis overestimated the performance resulting in more than 2\,Mbps difference at $N/M=1$\footnote{November 29, 2012: This discrepancy was only due to selection of $M_D=M$ for the model of~\cite{pawelczak_tvt_2009}, instead of $M_D=M-1$. See also Section~\ref{sec:BNS}.}. Interestingly, if we consider the same set of parameters as in Section~\ref{sec:pu_level_impact} then the model of~\cite{pawelczak_tvt_2009} almost agrees with the model of our paper. Since the set of parameters that has been chosen in~\ref{sec:pu_level_impact} are similar to~\cite{pawelczak_tvt_2009} we remark that the observations on the performance of this OSA MAC in~\cite{pawelczak_tvt_2009} were reflecting the reality.

\subsubsection{Impact of Channel Errors on the OSA Multichannel MAC Performance}
\label{sec:error_results}

\begin{figure}
\centering
\subfigure[]{\includegraphics[width=0.49\columnwidth]{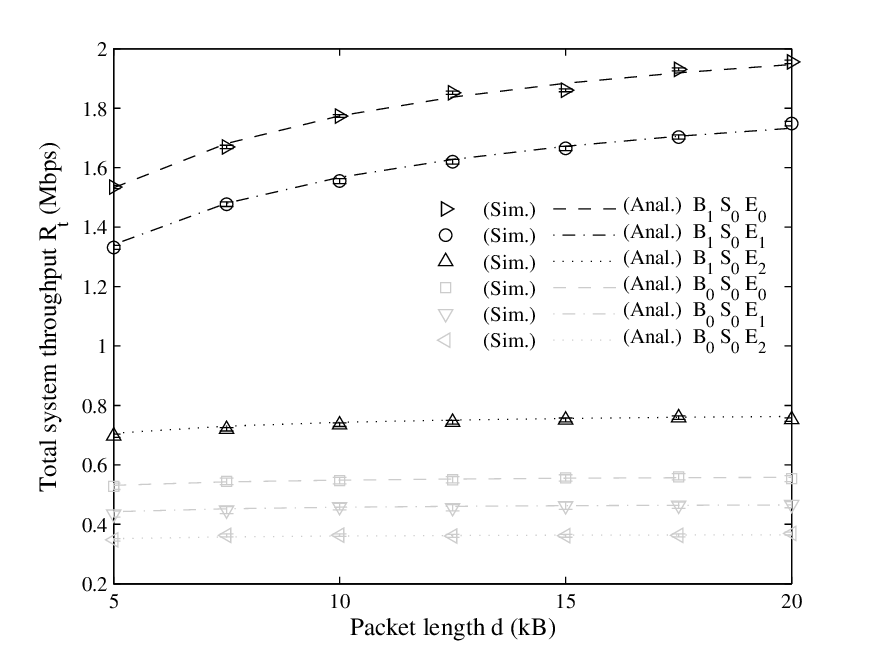}\label{fig:su_packet_impact1}}
\subfigure[]{\includegraphics[width=0.49\columnwidth]{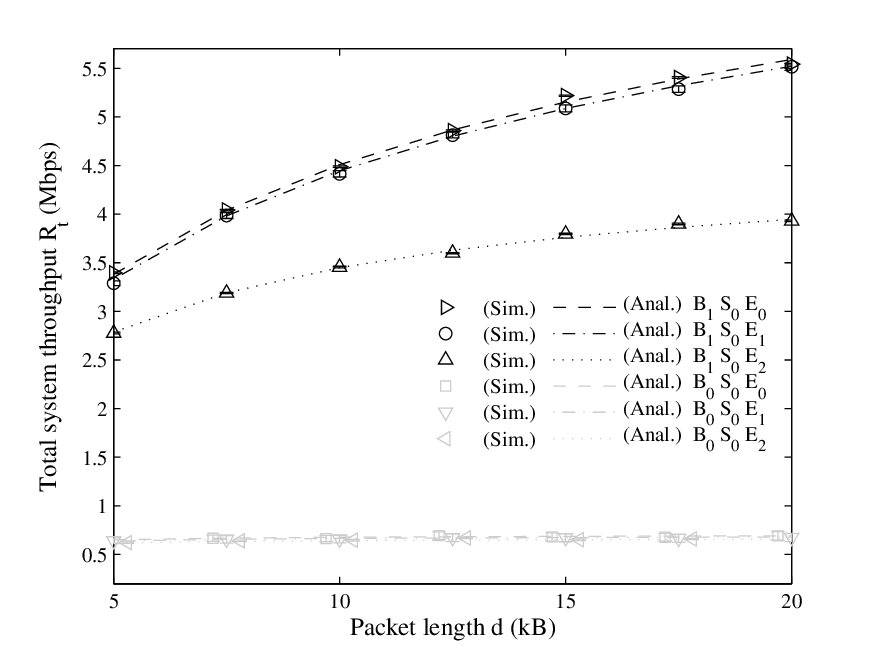}\label{fig:su_packet_impact2}}
\caption{Throughput of HCC OSA MAC as a function of packet size $d$ for (a) $M=3$, $N=12$, $p_e=0.1$ and (b) $M=12$, $N=40$, $p_e=0.01$ and two distinct error handling startegies. Rest of the parameters are the same as in Fig.~\ref{fig:q_p_impact}, except for $q_p=0.1$. E$_1$ and E$_2$ denote error models described in Section~\ref{sec:Error}. E$_0$ denotes the system with $p_e=0$.}
\label{fig:su_packet_impact_error}
\end{figure}

To observe the impact of channel errors on the MAC protocol throughput we have set up the following experiment. For HCC and both network sizes, small and large, we have observed the average throughput for different SU packet lengths and channel error probabilities. The results are presented in Fig.~\ref{fig:su_packet_impact_error}. For comparison in Fig.~\ref{fig:su_packet_impact_error} we present the system with no errors, denoted as E$_0$. We kept values of $p_e$ realistic, not exceeding 1\%. Obviously system with punctured errors $E_1$ obtains much higher throughput than system E$_2$, since more data can be potentially sent after one control packet exchange. Again, buffering allows to obtain higher throughput in comparison to non-buffered case, even with the data channel errors present. Note that system E$_2$ is more prone to errors than E$_1$, observe Fig.~\ref{fig:su_packet_impact1} and Fig.~\ref{fig:su_packet_impact2} for B$_1$S$_0$ E$_1$ and B$_1$S$_0$ E$_2$.

\subsubsection{Impact of PU Channel Occupancy Distributions on the OSA Multichannel MAC Performance}
\label{sec:distribution_impact}

All previous analyses were done under the assumption that traffic generated by SU and channel occupancy of PU can be described by the geometric process. This assumption holds generally either for SU traffic or PU channel occupancy statistics. For example, it has been shown recently in~\cite{wellens_phycom_2009} that geometric process constitutes more than 60\% of the measured PU traffic in GSM 900 uplink, GSM 1800 downlink, DECT and 2.4\,GHz UNII channels. It is important however to see the behavior of the considered data MAC protocols with other traffic distributions. Since the impact of different SU packet length distributions has been investigated in~\cite[Sec. 5.2]{mo_tmc_2008}, concluding that comparable throughput of multichannel MAC protocols is obtained, we focus on the impact of different PU traffic distributions on the OSA network performance. Due to vast number of combinations of protocol and traffic distributions we have narrowed our presentation to DCC and the following distributions: i) discrete uniform (denoted symbolically as U), ii) log-normal (denoted symbolically as L), and for comparison iii) geometric (denoted symbolically as E) used in the analysis. We have tested the protocol performance for different combinations of ``on'' and ``off'' times of PU activity. These were EE, LE, EL, LL (all possible combinations of ``on'' and ``off'' times obtained in~\cite[Tab. 3 and Tab. 4]{wellens_phycom_2009}) and additionally EU, UU, where first and second letter denotes selected distribution for ``on'' and ``off'' times, respectively. Due to the complexity of the analysis we show only the simulation results using the same simulation method of batch means, with the same parameters as described at the beginning of Section~\ref{sec:numerical_results}.

The parameter of each distribution was selected such that the mean value of each distribution was equal to $1/p_c$  for ``on'' time and $1-1/p_c$ for ``off'' time. The uniform distribution has a non-continuous set of mean values, $(a_b+a_n)/2$, where $a_b,a_n\in \mathbb{N}$ denoting lower and upper limit of the distribution, respectively, which precludes existence of every mean on or off value for $p_c\in(0,1)$. To solve that problem an continuous uniform distribution with required mean was used and rounded to the highest integer. This resulted in a slightly lower last peak in the probability mass function at $a_{n}$ for $1/p_{c}\notin \mathbb{N}$ or $1-1/p_{c}\notin \mathbb{N}$. In case of log-normal distribution, because it is continuous, it was rounded it to the nearest integer as well, with scale parameter $\sigma=\sqrt{\log\left(\frac{v_{l}}{c_{l}^2}+1\right)}$ and location parameter $\mu=\log\left(\frac{c_{l}^2}{\sqrt{v_{l}+c_{l}^2}}\right)$, where $c_l=1/p_{c}$, $v_l=(1-p_{c})/p_{c}^2$ is the mean and variance of the resulting discretized log-normal distribution. Note that the variance of the used discretized log-normal distribution is equal to the variance of geometric distribution for the same mean value. The variance of resulting discretized uniform continuos distribution could not be equal to the variance of the geometric distribution due the reasons described earlier.
\begin{figure}
\centering
\subfigure[DCC B$_{0}$S$_{0}$, SN]{\includegraphics[width=0.24\columnwidth]{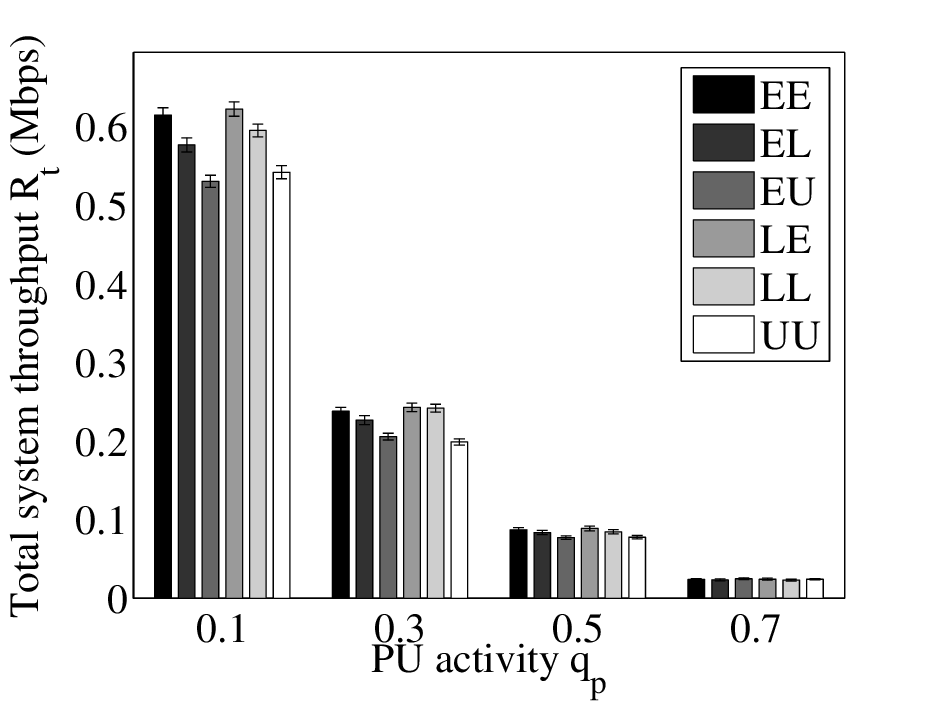}\label{fig:dist_B0S0_small}}
\subfigure[DCC B$_{0}$S$_{1}$, SN]{\includegraphics[width=0.24\columnwidth]{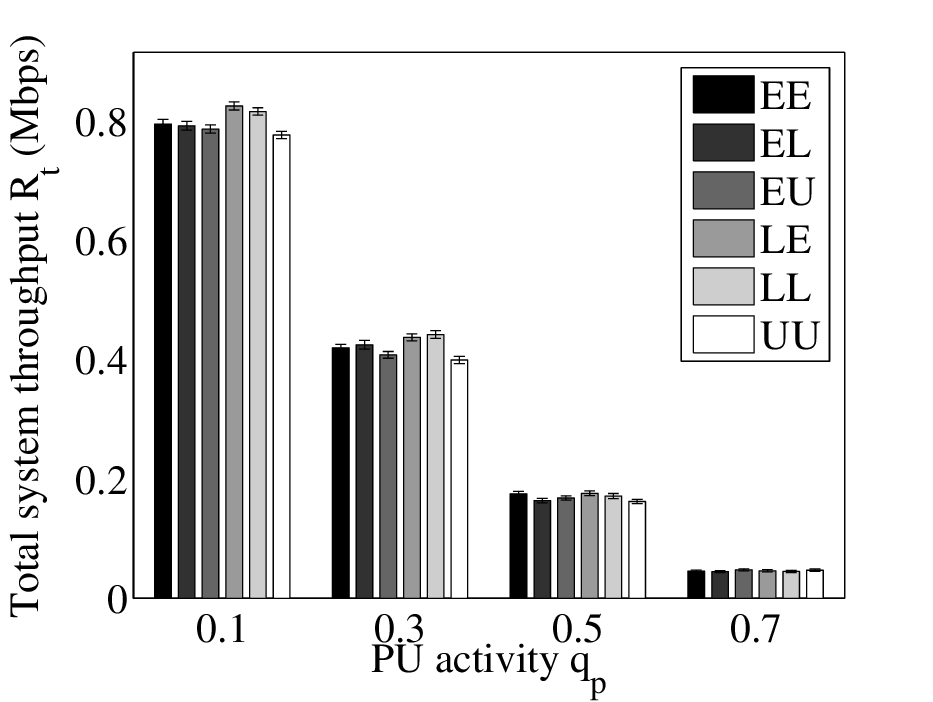}\label{fig:dist_B0S1_small}}
\subfigure[DCC B$_{1}$S$_{0}$, SN]{\includegraphics[width=0.24\columnwidth]{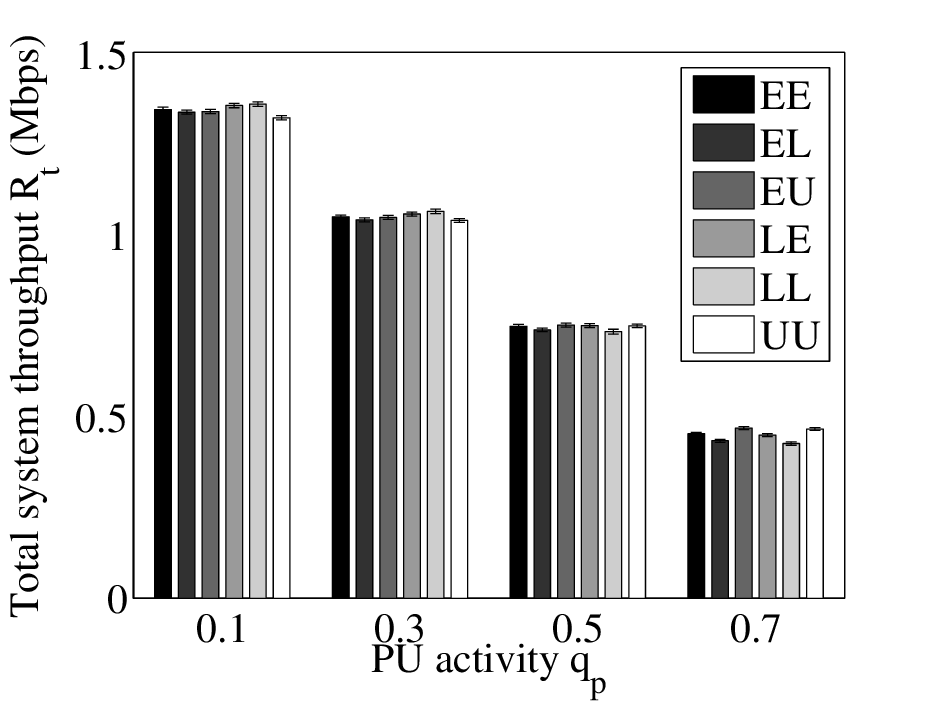}\label{fig:dist_B1S0_small}}
\subfigure[DCC B$_{1}$S$_{1}$, SN]{\includegraphics[width=0.24\columnwidth]{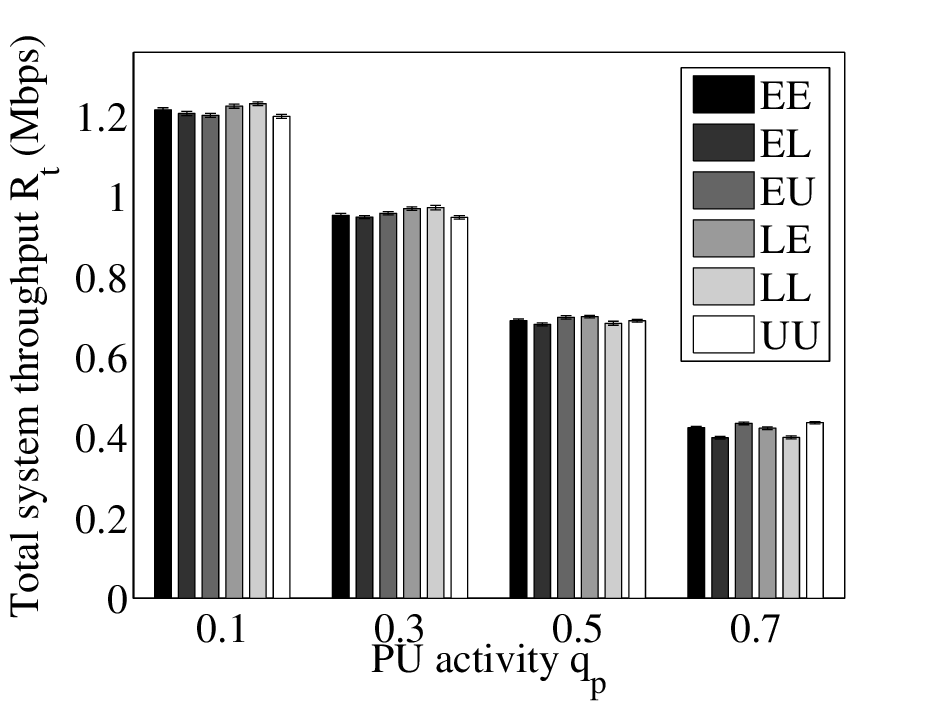}\label{fig:dist_B1S1_small}}
\subfigure[DCC B$_{0}$S$_{0}$, LN]{\includegraphics[width=0.24\columnwidth]{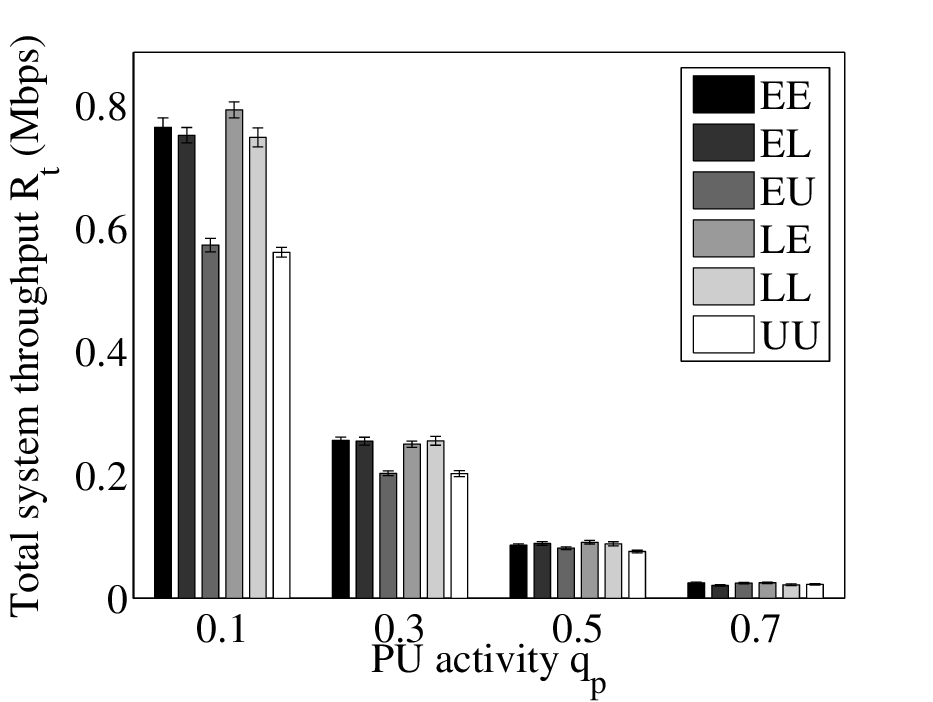}\label{fig:dist_B0S0_large}}
\subfigure[DCC B$_{0}$S$_{1}$, LN]{\includegraphics[width=0.24\columnwidth]{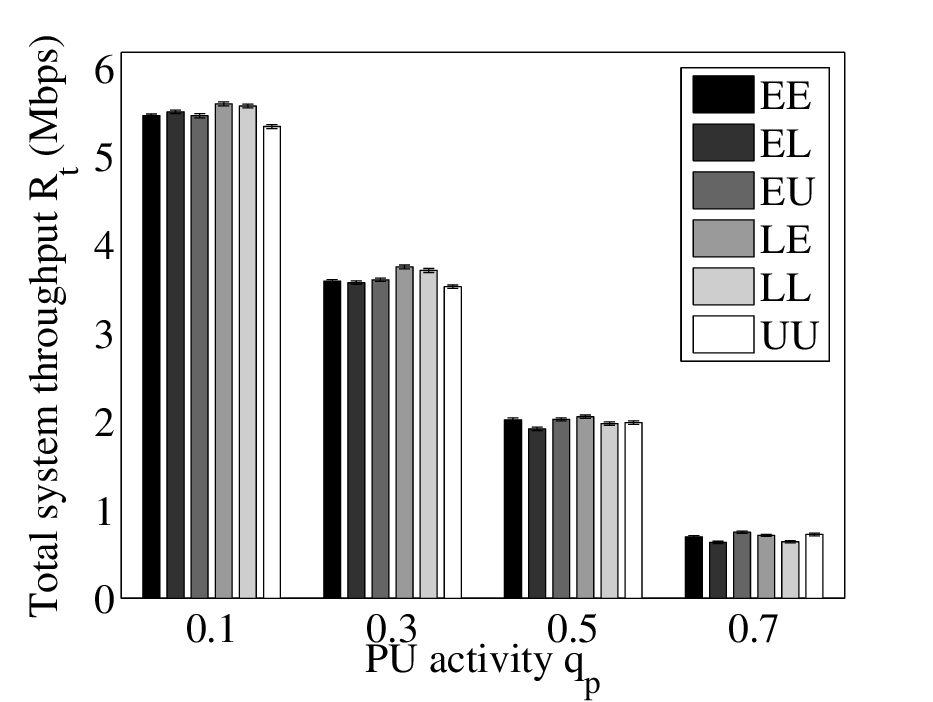}\label{fig:dist_B0S1_large}}
\subfigure[DCC B$_{1}$S$_{0}$, LN]{\includegraphics[width=0.24\columnwidth]{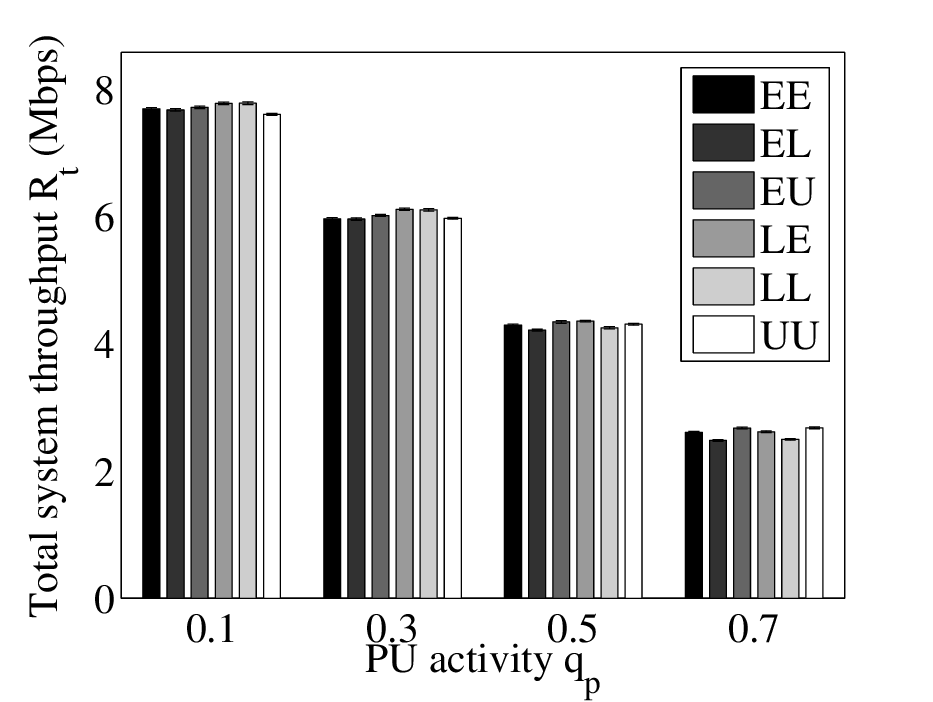}\label{fig:dist_B1S0_large}}
\subfigure[DCC B$_{1}$S$_{1}$, LN]{\includegraphics[width=0.24\columnwidth]{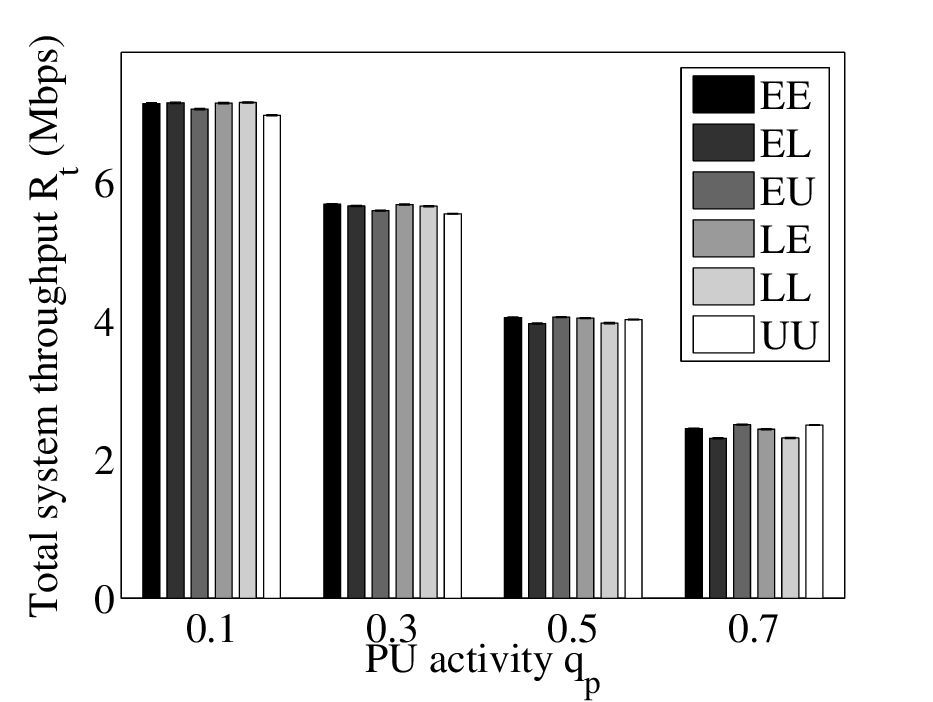}\label{fig:dist_B1S1_large}}
\caption{Impact of different PU on and off times distributions on OSA DCC multichannel MAC performance; (a)--(d) small scale network, (e)--(h) large scale network, as described in as in Fig.~\ref{fig:q_p_impact}. E, U, L denote geometric, uniform and log-normal distribution respectively, where the first and second parameter in the legend denotes off and on time, respectively. SN denotes small network and LN denotes large network.}
\label{fig:pu_distribution_impact}
\end{figure}

The results are presented in Fig.~\ref{fig:pu_distribution_impact}. We focus on two network types, as indicated earlier: (i) large scale and (ii) small scale, with the assumed parameters as in Fig.~\ref{fig:q_p_impact}. We select four values of $q_p$ for the clarity of the presentation. The most important observation is that irrespective of the considered distribution DCC obtains relatively the same throughput and the same relation between different protocol options exists as it was shown analytically in Fig.~\ref{fig:q_p_impact}. If one wants to select the distribution combination with the highest throughput it would be LE and LL, while the throughput obtained being almost equal to the one obtained via analysis for the geometric distribution. The distribution with the lowest throughput is UU and EU, due to the difference of the second moment between the other two distributions for the on time. The difference in throughput between UU, EU and the remaining distributions is more visible for the large network. The most surprising result of this investigation is that any DCC MAC protocol option with buffering removes the impact of distribution type on the obtained performance, compare Fig.~\ref{fig:dist_B0S0_small} and Fig.~\ref{fig:dist_B1S0_small}, or Fig.~\ref{fig:dist_B0S0_large} with Fig.~\ref{fig:dist_B1S1_large} for any value of $q_p$.

\subsection{Performance of Joint Spectrum Sensing and OSA MAC Protocols}
\label{sec:joint_performance}

Having results for spectrum sensing protocol and OSA MAC we join these two layers to form a complete OSA network stack. By means of exhaustive search we solve the optimization problem of (\ref{eq:framework}). We will also investigate the set of parameters that maximize $R_t$ for small and large scale network. 

We divide our analysis in macroscopic and microscopic case observing $R_t$ for small scale network with $M=3$, $N=12$, $d=5$\,kB, and large scale network with $M=12$, $N=40$, $d=20$\,kB. For each case we select a set of spectrum sensing and OSA MAC protocols that are possible and, as we believe, most important to the research community. For a fixed set of parameters $C=1$\,Mbps, $b=1$\,MHz, $p=e^{-1}/N$, $t_{d,\max}=1$\,ms (microscopic case), $t_{d,\max}=2$\,s (macroscopic case), $\alpha=1/M$, $t_t=1$\,ms, $p_{d,\min}=0.99$, $\gamma=-5$\,dB, $q_p=0.1$, and $t_p=100$\,$\mu$s we leave $\kappa$, $t_e$, $n_g$, and $p_f$ as optimization variables.

\subsubsection{Microscopic Model}
\label{sec:macroscopic_case_results}

\begin{figure}
\subfigure[]{\includegraphics[width=0.49\columnwidth]{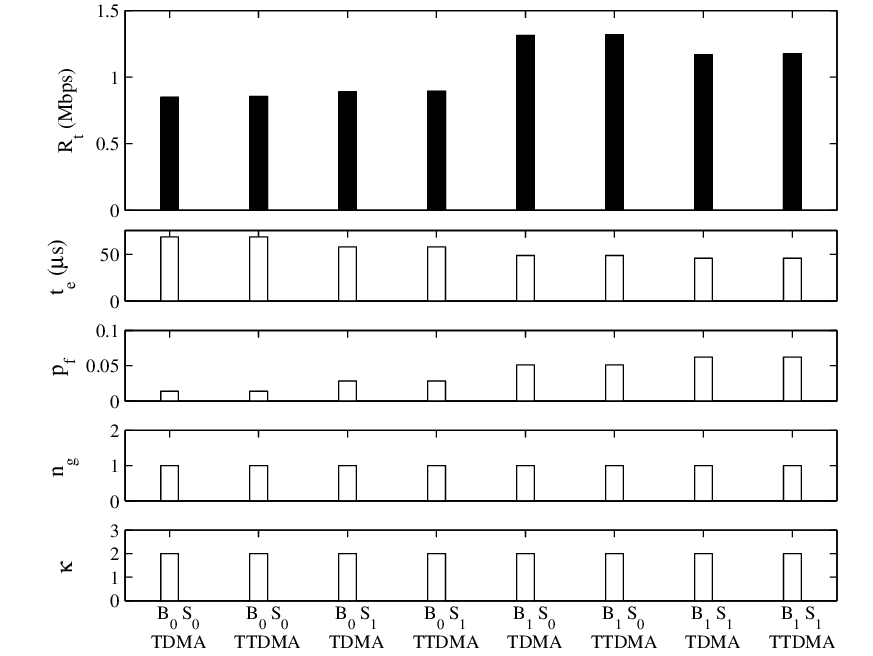}\label{fig:M3N12_micro}}
\subfigure[]{\includegraphics[width=0.49\columnwidth]{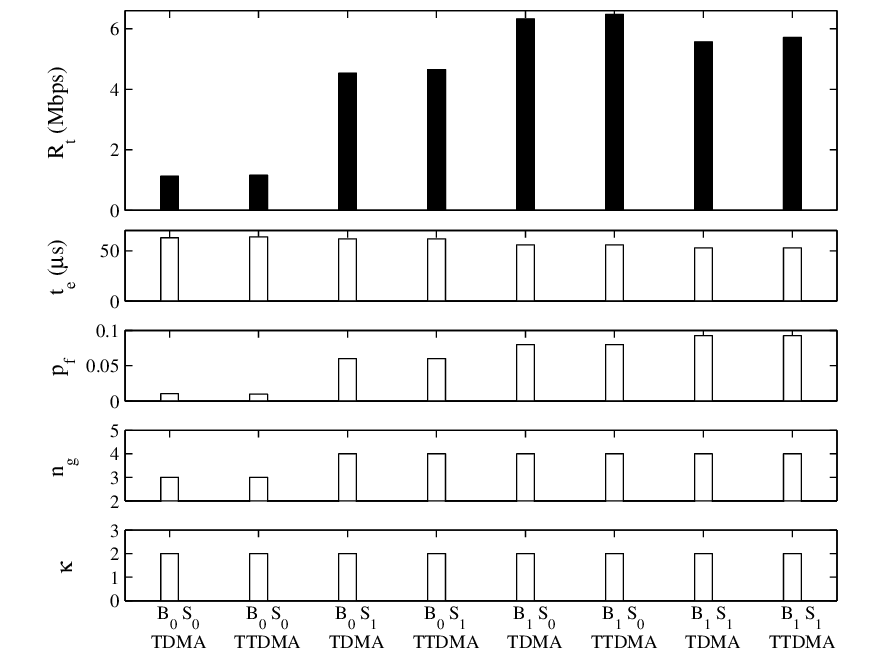}\label{fig:M12N40_micro}}
\caption{Optimization result of the selected protocol combination with DCC for the microscopic model for (a) $M=3$, $N=12$, and (b) $M=12$, $N=40$. Common parameters: $p_e=0$, $d=5$\,kB, $C=1$\,Mbps, $b=1$\,MHz, $p=e^{-1}/N$, $t_{d,\max}=1$\,ms, $\alpha=1/M$, $t_t=1$\,ms, $p_{d,\min}=0.99$, $\gamma=-5$\,dB, $q_p=0.1$, and $t_p=100$\,ms.}
\label{fig:micro_opt}
\end{figure}

Here we focus only on DCC protocol, since collaborative spectrum sensing is only possible via a PU free control channel, which is inefficient to accomplish with HCC. Also, for sensing measurement dissemination we do not consider SSMA, which would be most difficult to implement in practice. The results are presented in Fig.~\ref{fig:micro_opt}.

DCC B$_1$S$_0$ with TTDMA is the best option, both for small scale and large scale network, see Fig.~\ref{fig:M3N12_micro} and Fig.~\ref{fig:M12N40_micro}, respectively. Because of relatively high switching time B$_1$S$_1$ performs slightly worse than B$_1$S$_0$, for small and large scale network. DCC B$_0$S$_0$ with TDMA is the worst protocol combination, which confirms earlier results from Section~\ref{sec:spectrum_sensing_performance} and Section~\ref{sec:mac_performance}. Irrespective of network size it is always better to buffer SU connections preempted by PU than to look for vacant channels, compare again B$_1$S$_0$ and B$_0$S$_1$ in Fig.~\ref{fig:M3N12_micro} and Fig.~\ref{fig:M12N40_micro}. The difference between B$_0$S$_0$ and B$_0$S$_1$ is mostly visible for a large network scenario, see Fig.~\ref{fig:M12N40_micro}, since with a large number of channels there are more possibilities to look for empty channels. 

For all protocol combinations and both network sizes $\kappa=2$ maximizes throughput performance, see Fig.~\ref{fig:M3N12_micro}. Interestingly, network size dictates the size of a sensing group. For small scale network, $n_g=1$ is the optimal value, see Fig.~\ref{fig:M3N12_micro}, but for a large network $R_t$ is maximized when $n_g=3$ (for B$_0$S$_0$) and $n_g=4$ (for the rest). We can conclude that with a small network it is better to involve all nodes in sensing, while for larger networks it is better to divide them into groups, which agrees with the observation from Section~\ref{sec:group_selection}. Moreover, we observe that the performance difference between TTDMA and TDMA is not as big as in Fig.~\ref{fig:sensing_gr1} when parameters are optimized.

The most interesting result is observed for $p_f$. With the increase of protocol complexity false alarm increases as well. Also with an increase of $p_f$, quiet time is decreasing. Because buffering and switching improves the performance, there can be more margin to design the spectrum sensing.

\subsubsection{Macroscopic Model}
\label{sec:microscopic_case_results}

\begin{figure}
\subfigure[]{\includegraphics[width=0.49\columnwidth]{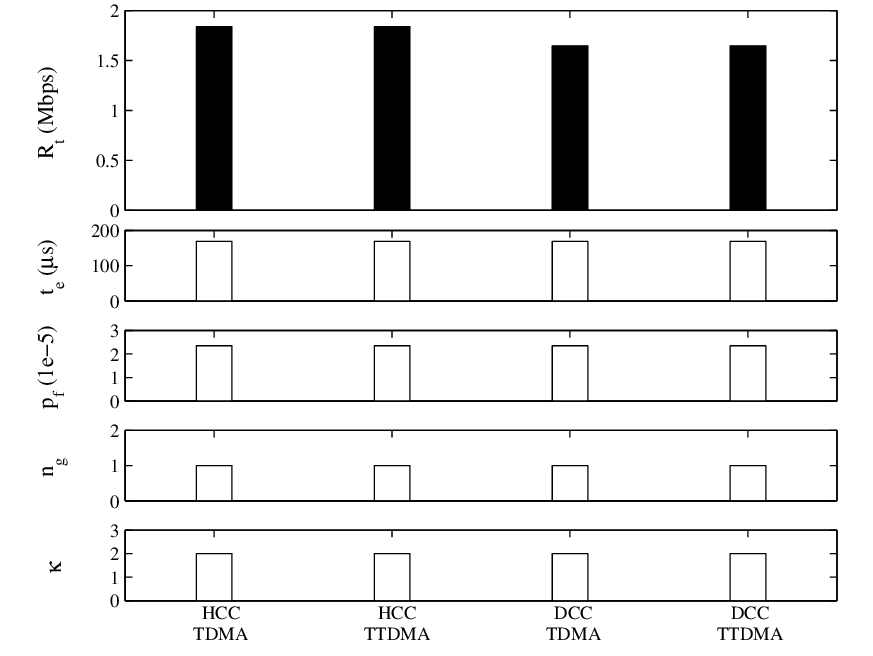}\label{fig:M3N12_macro}}
\subfigure[]{\includegraphics[width=0.49\columnwidth]{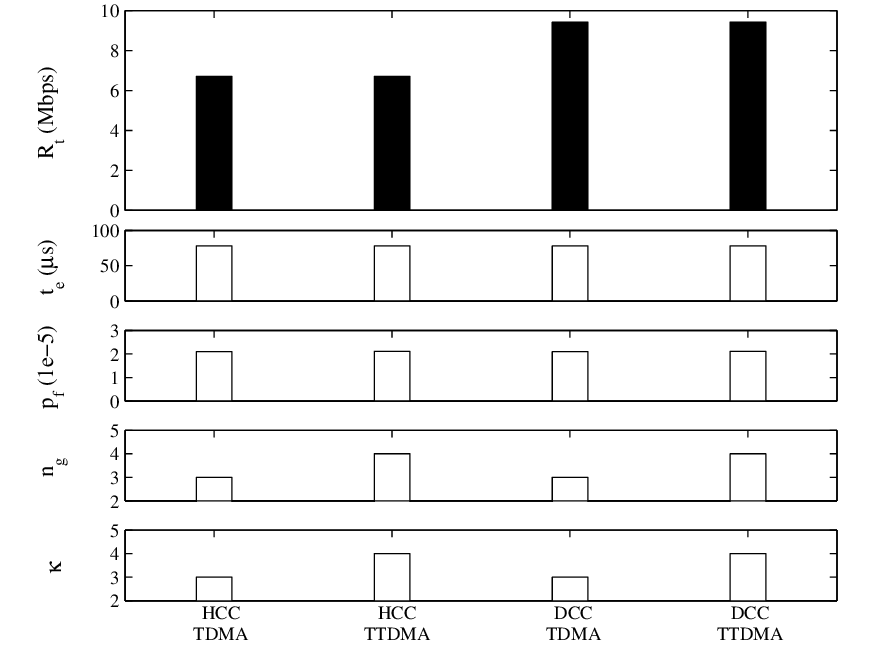}\label{fig:M12N40_macro}}
\caption{Optimization result of the selected protocol combination for the macroscopic model for (a) $M=3$, $N=12$, and (b) $M=12$, $N=40$. Common parameters are the same as in Fig.~\ref{fig:micro_opt}, except for $t_{d,\max}=2$\,s.}
\label{fig:macro_opt}
\end{figure}

For the macroscopic model we explore both non-OSA DCC and HCC with TDMA and TTDMA as sensing protocols. The results are presented in Fig.~\ref{fig:macro_opt}.

DCC obtains higher throughput than HCC for a small scale network, and vice versa, compare Fig.~\ref{fig:M3N12_macro} and Fig.~\ref{fig:M12N40_macro}, respectively. This confirms the observations of~\cite[Fig. 3]{pawelczak_tvt_2009},~\cite[Fig. 3]{mo_tmc_2008}. Just like in Fig.~\ref{fig:M3N12_micro}, for small scale network $\kappa=2$ and $n_g=2$ are the ones that maximize $R_t$. For the large scale network, however, $\kappa=3$ and $n_g=3$ is optimal for TDMA, and $\kappa=4$ and $n_g=4$ for TTDMA. This means that for large networks it is beneficial to split the network into smaller groups. Again, this confirms our findings from Section~\ref{sec:macroscopic_case_results}. For both network scenarios $p_f$ and $t_e$ is relatively the same for all protocols considered.

Note that for the large scale network in the macroscopic model, an SU takes more time to detect a PU than in the microscopic model because large $t_{d,max}$ reduces the time overhead. The release of time restriction impacts the large scale network by requiring greater value of $\kappa$ to achieve the maximum throughput.

\section{Conclusion}
\label{sec:conclusions}

We have presented a comprehensive framework enabling assessment of the performance of joint spectrum sensing and MAC protocol operation for OSA networks. In the model we have proposed we focused on the link layer throughput as the fundamental metric to assess performance. We have parameterized spectrum sensing architectures for energy detection based systems with collaborative measurements combining. We have proposed a novel spectrum sensing MAC denoted Truncated Time Division Multiple Access. We have also categorized multichannel MAC protocols for OSA networks based on their ability to buffer and switch existing SU connections on the arrival of a PU. Our analysis is supported by simulations which prove the accuracy of the obtained expressions.

Some of the design guidelines that need to be noted are as follows. For spectrum sensing introducing TTDMA gives an improvement in obtained performance in compared to TDMA. Large networks, i.e. having many channels and users, benefit from clustering, while for small networks it is better to create small number of clusters such that sensing time is optimized. When considering MAC protocol design for OSA it is clear that more benefit comes from introducing SU connection buffering than channel switching, for those SU connections that have been preempted by PU. Interestingly, although intuition would suggest that MAC protocols that combine SU connection buffering and channel switching would outperform all other protocols, due to switching overhead this combination is usually inferior to protocols that involve only SU connection buffering.

Our future task will be to investigate the delay experience by using any of OSA MAC protocols proposed. We plan to develop a comprehensive simulation software which will implement features not covered by our model, like queue per each SU.


\begin{thebibliography}{10}
\providecommand{\url}[1]{#1}
\csname url@rmstyle\endcsname
\providecommand{\newblock}{\relax}
\providecommand{\bibinfo}[2]{#2}
\providecommand\BIBentrySTDinterwordspacing{\spaceskip=0pt\relax}
\providecommand\BIBentryALTinterwordstretchfactor{4}
\providecommand\BIBentryALTinterwordspacing{\spaceskip=\fontdimen2\font plus
\BIBentryALTinterwordstretchfactor\fontdimen3\font minus
  \fontdimen4\font\relax}
\providecommand\BIBforeignlanguage[2]{{%
\expandafter\ifx\csname l@#1\endcsname\relax
\else
\language=\csname l@#1\endcsname
\fi
#2}}

\bibitem{park_dyspan_submitted}
J.~{Park}, P.~{Pawe{\l}czak}, and D.~\v{C}abri\'{c}, ``To buffer or to switch:
  Design of multichannel {MAC} for {OSA} ad hoc networks,'' in \emph{Proc. IEEE
  DySPAN}, Singapore, Apr. 6--9, 2010.

\bibitem{staple_spectrum_2004}
G.~{Staple} and K.~{Werbach}, ``The end of spectrum scarcity,'' \emph{{IEEE}
  Spectr.}, vol.~41, no.~3, pp. 48--52, Mar. 2004.

\bibitem{Prasad_commag_2007}
R.~V. {Prasad}, P.~Pawe{\l}czak, J.~{Hoffmeyer}, and S.~{Berger}, ``Cognitive
  functionality in next generation wireless networks: Standardization
  efforts,'' \emph{{IEEE} Commun. Mag.}, vol.~46, no.~4, pp. 72--78, Apr. 2008.

\bibitem{noam_commag_1995}
E.~Noam, ``Taking the next step beyond spectrum auctions: Open spectrum
  access,'' \emph{{IEEE} Commun. Mag.}, vol.~33, no.~12, pp. 66--73, Dec. 1995.

\bibitem{Zhao_sigprocmag_2007}
Q.~{Zhao} and B.~M. {Sadler}, ``A survey of dynamic spectrum access: Signal
  processing, networking, and regulatory policy,'' \emph{{IEEE} Signal
  Processing Mag.}, vol.~24, no.~3, pp. 79--89, May 2007.

\bibitem{hoffmeyer_scc41_2008}
J.~{Hoffmeyer}, D.~{Stewart}, S.~{Berger}, B.~{Eydt}, F.~{Frantz},
  F.~{Granelli}, K.~{Kontson}, D.~{Murotake}, K.~{Nolan}, P.~{Pawe{\l}czak},
  R.~V. {Prasad}, R.~{Roy}, M.~{Scoville}, D.~{Sicker}, D.~{Swain}, and
  P.~{Tenhula}, \emph{Definitions and Concepts for Dynamic Spectrum Access:
  Terminology Relating to Emerging Wireless Networks, System Functionality, and
  Spectrum Management}, IEEE 1900.1-2008 Standard, Oct. 2, 2008.

\bibitem{park_icc_2009}
J.~{Park}, R.~{Jain}, and D.~\v{C}abri\'{c}, ``Spectrum sensing design
  framework based on cross-layer optimization of detection efficiency,'' in
  \emph{Proc. IEEE ICC}, Dresden, Germany, EU, June 14--18, 2009.

\bibitem{timmers_tvt_2009}
M.~{Timmers}, S.~{Pollin}, A.~{Dejonge}, L.~{Van der Perre}, and F.~{Catthoor},
  ``A distributed multichannel {MAC} protocol for multihop cognitive radio
  networks,'' \emph{{IEEE} Trans. Veh. Technol.}, vol.~59, no.~1, pp. 446--459,
  Jan. 2010.

\bibitem{Jia_jsac_2008}
J.~{Jia}, Q.~{Zhang}, and X.~{Shen}, ``{HC-MAC}: A hardware-constrained
  cognitive {MAC} for efficient spectrum management,'' \emph{{IEEE} J. Select.
  Areas Commun.}, vol.~26, no.~1, pp. 106--117, Jan. 2008.

\bibitem{stevenson_commag09}
C.~R. {Stevenson}, G.~{Chouinard}, Z.~{Lei}, W.~{Hu}, S.~J. {Shellhammer}, and
  W.~{Caldwell}, ``{IEEE 802.22}: The first cognitive radio wireless regional
  area network standard,'' \emph{{IEEE} Commun. Mag.}, vol.~47, no.~1, pp.
  130--138, Jan. 2009.

\bibitem{cordeiro_book09}
C.~{Cordeiro}, D.~{Cavalcanti}, and S.~{Nandagopalan}, ``Cognitive radio for
  broadband wireless access in {TV} bands: The {IEEE} 802.22 standard,'' in
  \emph{Cognitive Radio Communications and Networks: Principles and Practice},
  A.~M. {Wyglinski}, M.~{Nekovee}, and Y.~T. {Hou}, Eds.\hskip 1em plus 0.5em
  minus 0.4em\relax Amsterdam, The Netherlands: Elsevier, Inc., 2009.

\bibitem{Liang_twc_2008}
Y.-C. {Liang}, Y.~{Zeng}, E.~C. {Peh}, and A.~T. {Hoang}, ``Sensing throughput
  tradeoff in cognitive radio networks,'' \emph{{IEEE} Trans. Wireless
  Commun.}, vol.~7, no.~4, pp. 1326--1337, Apr. 2008.

\bibitem{peh_tvt_2009}
E.~C.~Y. {Peh}, Y.-C. {Liang}, Y.~L. {Guan}, and Y.~{Zeng}, ``Optimization of
  cooperative sensing in cognitive radio networks: A sensing-throughput
  tradeoff view,'' \emph{{IEEE} Trans. Veh. Technol.}, vol.~58, no.~9, pp.
  5294--5299, Nov. 2009.

\bibitem{jeon_twc_2008}
W.~S. {Jeon}, D.~G. {Jeong}, J.~A. {Han}, G.~{Ko}, and M.~S. {Song}, ``An
  efficient quiet period management scheme for cognitive radio systems,''
  \emph{{IEEE} Trans. Wireless Commun.}, vol.~7, no.~2, pp. 505--509, Feb.
  2008.

\bibitem{pawelczak_tvt_2009}
P.~Pawe{\l}czak, S.~{Pollin}, H.-S.~W. {So}, A.~{Bahai}, R.~V. {Prasad}, and
  R.~{Hekmat}, ``Performance analysis of multichannel medium access control
  algorithms for opportunistic spectrum access,'' \emph{{IEEE} Trans. Veh.
  Technol.}, vol.~58, no.~6, pp. 3014--3031, July 2009.

\bibitem{Papadimitratos_commag_2005}
P.~{Papadimitratos}, S.~{Sankaranarayanan}, and A.~{Mishra}, ``A bandwidth
  sharing approach to improve licensed spectrum utilization,'' \emph{{IEEE}
  Commun. Mag.}, vol.~43, no.~12, pp. S10--S14, Dec. 2005.

\bibitem{Hoang_twc_2009}
A.~T. {Hoang}, Y.-C. {Liang}, D.~T.~C. {Wong}, Y.~{Zeng}, and R.~{Zhang},
  ``Opportunistic spectrum access for energy-constrained cognitive radios,''
  \emph{{IEEE} Trans. Wireless Commun.}, vol.~8, no.~3, pp. 1206--1211, Mar.
  2009.

\bibitem{wang_tmc_2009}
X.~Y. {Wang}, A.~{Wong}, and P.-H. {Ho}, ``Extended knowledge-based reasoning
  approach to spectrum sensing for cognitive radio,'' \emph{{IEEE} Trans.
  Mobile Comput.}, vol.~9, no.~4, pp. 465--478, Apr. 2010.

\bibitem{hossain_book_2009}
E.~{Hossain}, D.~{Niyato}, and Z.~{Han}, \emph{Dynamic Spectrum Access and
  Management in Cognitive Radio Networks}.\hskip 1em plus 0.5em minus
  0.4em\relax New York, NY, USA: Cambridge University Press, 2009.

\bibitem{Chou_jsac_2007}
C.-T. {Chou}, S.~{Shankar N}, H.~{Kim}, and K.~G. {Shin}, ``What and how much
  to gain by spectrum agility?'' \emph{{IEEE} J. Select. Areas Commun.},
  vol.~25, no.~3, pp. 576--588, Apr. 2007.

\bibitem{srinivasa_twc_2008}
S.~{Srinivasa} and S.~A. {Jafar}, ``How much spectrum sharing is optimal in
  cognitive radio networks?'' \emph{{IEEE} Trans. Wireless Commun.}, vol.~7,
  no.~10, pp. 4010--4018, Oct. 2008.

\bibitem{tang_twc_2008}
S.~{Tang} and B.~L. {Mark}, ``Modelling and analysis of opportunistic spectrum
  sharing with unreliable spectrum sensing,'' \emph{{IEEE} Trans. Wireless
  Commun.}, vol.~8, no.~4, pp. 1934--1943, Apr. 2009.

\bibitem{tang_twc_2009}
------, ``Analysis of oportunistic spectrum sharing with {Markovian} arrivals
  and phase-type service,'' \emph{{IEEE} Trans. Wireless Commun.}, vol.~8,
  no.~6, pp. 3142--3150, June 2009.

\bibitem{kalil_asmta_2009}
M.~A. {Kalil}, H.~{Al-Mahdi}, and A.~{Mitschele-Thiel}, ``Analysis of
  opportunistic spectrum access in cognitive ad hoc networks,'' in \emph{Proc.
  ASMTA}, Madrid, Spain, EU, June 9--12, 2009.

\bibitem{zhu_commlett_2007}
X.~{Zhu}, L.~{Shen}, and T.-S.~P. {Yum}, ``Analysis of cognitive radio spectrum
  access with optimal channel reservation,'' \emph{{IEEE} Commun. Lett.},
  vol.~11, no.~4, pp. 304--306, Apr. 2007.

\bibitem{wong_commlett_2009}
E.~W.~M. {Wong} and C.~H. {Foh}, ``Analysis of cognitive radio spectrum access
  with finite user population,'' \emph{{IEEE} Commun. Lett.}, vol.~13, no.~5,
  pp. 294--296, May 2009.

\bibitem{zhangyan_icc_2008}
Y.~{Zhang}, ``Dynamic spectrum access in cognitive radio wireless networks,''
  in \emph{Proc. IEEE ICC}, Beijing, China, May 19--23, 2008.

\bibitem{wong_wcnc_2008}
D.~T.~C. {Wong}, A.~T. {Hoang}, Y.-C. {Liang}, and F.~P.~S. {Chin}, ``Dynamic
  spectrum access with imperfect sensing in open spectrum wireless networks,''
  in \emph{Proc. IEEE WCNC}, Las Vegas, NV, USA, Mar. 31~--~Apr. 3, 2008.

\bibitem{wang_twc_2009}
B.~{Wang}, Z.~{Li}, K.~J.~R. {Liu}, and T.~C. {Clancy}, ``Primary-prioritized
  {Markov} approach for dynamic spectrum allocation,'' \emph{{IEEE} Trans.
  Wireless Commun.}, vol.~8, no.~4, pp. 1854--1865, Apr. 2009.

\bibitem{lee_commlett_2009}
H.~{Lee} and D.-H. {Cho}, ``{VoIP} capacity analysis in cognitive radio
  system,'' \emph{{IEEE} Commun. Lett.}, vol.~13, no.~6, pp. 393--395, June
  2009.

\bibitem{luo_twc_2009}
L.~{Luo}, N.~M. {Neihart}, S.~{Roy}, and D.~J. {Allstot}, ``A two-stage sensing
  technique for dynamic spectrum access,'' \emph{{IEEE} Trans. Wireless
  Commun.}, vol.~8, no.~6, pp. 3028--3037, June 2009.

\bibitem{hamdaoui_twc_2009}
B.~{Hamdaoui}, ``Adaptive spectrum assessment for opportunistic access in
  cognitive radio networks,'' \emph{{IEEE} Trans. Wireless Commun.}, vol.~8,
  no.~2, pp. 922--930, Feb. 2009.

\bibitem{lee_twc_2008}
W.-Y. {Lee} and I.~F. {Akyildiz}, ``Optimal spectrum sensing framework for
  cognitive radio networks,'' \emph{{IEEE} Trans. Wireless Commun.}, vol.~7,
  no.~10, pp. 3845--3857, Oct. 2008.

\bibitem{mo_tmc_2008}
J.~{Mo}, H.-S.~W. {So}, and J.~{Walrand}, ``Comparison of multichannel {MAC}
  protocols,'' \emph{{IEEE} Trans. Mobile Comput.}, vol.~7, no.~1, pp. 50--65,
  Jan. 2008.

\bibitem{Lai_arxiv_2007}
\BIBentryALTinterwordspacing
L.~{Lai}, H.~{El Gamal}, H.~{Jiang}, and H.~V. {Poor}. (2007) Cognitive medium
  access: Exploration, exploitation and competition. [Online]. Available:
  \url{http://arxiv.org/abs/0710.1385}
\BIBentrySTDinterwordspacing

\bibitem{Huang_tmc_2009}
S.~{Huang}, X.~{Liu}, and Z.~{Ding}, ``Optimal transmission strategies for
  dynamic spectrum access in cognitive radio networks,'' \emph{{IEEE} Trans.
  Mobile Comput.}, vol.~8, no.~12, pp. 1636--1648, Dec. 2009.

\bibitem{Kang_tvt_2009}
X.~{Kang}, Y.-C. {Liang}, H.~K. {Garg}, and L.~{Zhang}, ``Sensing-based
  spectrum sharing in cognitive radio networks,'' \emph{{IEEE} Trans. Veh.
  Technol.}, vol.~58, no.~8, pp. 4649--4654, Oct. 2009.

\bibitem{gronsund_pimrc_2009}
P.~{Gr{\o}nsund}, H.~N. {Pham}, and P.~E. {Engelstad}, ``Towards dynamic
  spectrum access in primary {OFDMA} systems,'' in \emph{Proc. IEEE PIMRC},
  Tokyo, Japan, Sept. 13--16, 2009.

\bibitem{Gerihofer_commag_2007}
S.~{Geirhofer}, L.~{Tong}, and B.~M. {Sandler}, ``Dynamic spectrum access in
  the time domain: Modeling and exploiting white space,'' \emph{{IEEE} Commun.
  Mag.}, vol.~45, no.~5, pp. 66--72, May 2007.

\bibitem{Huang_infocom_2009}
S.~{Huang}, X.~{Liu}, and Z.~{Ding}, ``Optimal sensing-transmission structure
  for dynamic spectrum access,'' in \emph{Proc. IEEE INFOCOM}, Rio De Janeiro,
  Brazil, Apr. 19--25, 2009.

\bibitem{gambini_twc_2008}
J.~{Gambini}, O.~{Simeone}, Y.~{Bar-Ness}, U.~{Spagnolini}, and T.~{Yu},
  ``Packet-wise vertical handover for unlicensed multi-standard spectrum access
  with cognitive radios,'' \emph{{IEEE} Trans. Wireless Commun.}, vol.~7,
  no.~12, pp. 5172--5176, Dec. 2008.

\bibitem{visser_vtc_2008}
F.~E. {Visser}, G.~J.~M. {Janssen}, and P.~{Pawe{\l}czak}, ``Multinode spectrum
  sensing based on energy detection for dynamic spectrum access,'' in
  \emph{Proc. IEEE VTC-Spring}, Singapore, May 11--14, 2008.

\bibitem{Zhang_ieeewc_2008}
W.~{Zhang} and K.~B. {Letaief}, ``Cooperative spectrum sensing with transmit
  and relay diversity in cognitive radio networks,'' \emph{{IEEE} Trans.
  Wireless Commun.}, vol.~7, no.~12, pp. 4761--4766, Dec. 2008.

\bibitem{Letaief_ieeeproc_2009}
K.~B. {Letaief} and W.~{Zhang}, ``Cooperative communications for cognitive
  radio networks,'' \emph{Proc. {IEEE}}, vol.~97, no.~5, pp. 878--893, May
  2009.

\bibitem{zhang_itw_2010}
W.~{Zhang}, A.~K. {Sadek}, C.~{Shen}, and S.~J. {Shellhammer}, ``Adaptive
  spectrum sensing,'' in \emph{Proc. Information Theory and Applications
  Workshop}, San Diego, CA, USA, Jan. 31~--~Feb. 5, 2010.

\bibitem{jiang_twc_2009}
H.~{Jiang}, L.~{Lai}, R.~{Fan}, and H.~V. {Poor}, ``Optimal selection of
  channel sensing order in cognitive radio,'' \emph{{IEEE} Trans. Wireless
  Commun.}, vol.~8, no.~1, pp. 297--307, Jan. 2009.

\bibitem{Li_ieeeglobecom_2009}
H.~{Li}, H.~{Dai}, and C.~{Li}, ``Collaborative quickest spectrum sensing via
  random broadcast in cognitive radio systems,'' in \emph{Proc. IEEE GLOBECOM},
  Hololulu, HI, USA, Nov. 30~--~Dec. 4, 2009.

\bibitem{tandra_procieee_2009}
R.~{Tandra}, S.~M. {Mishra}, and A.~{Sahai}, ``What is a spectrum hole and what
  does it take to recognize one?'' \emph{Proc. {IEEE}}, vol.~97, no.~5, pp.
  824--848, May 2009.

\bibitem{yucek_commsurv_2007}
T.~{Yucek} and H.~{Arslan}, ``A survey of spectrum sensing algorithms for
  cognitive radio applications,'' \emph{IEEE Comm. Surv. and Tut.}, vol.~11,
  no.~1, pp. 116--13, Jan. 2009.

\bibitem{biswas_icc_2009}
A.~R. {Biswas}, T.~C. {Aysal}, S.~{Kandeepan}, D.~{Kilazovich}, and
  R.~{Piesiewicz}, ``Cooperative shared spectrum sensing for dynamic cognitive
  radio network,'' in \emph{Proc. IEEE ICC}, Dresden, Germany, EU, June 14--18,
  2009.

\bibitem{sun_icc_2007}
C.~{Sun}, W.~{Zhang}, and K.~B. {Letaief}, ``Cluster-based cooperative spectrum
  sensing in cognitive radio systems,'' in \emph{Proc. IEEE ICC}, Glasgow,
  Scotland, EU, June 24--28, 2007.

\bibitem{weiss_scvt_2003}
T.~{Weiss}, J.~{Hillenbrand}, A.~{Krohn}, and F.~K. {Jondral}, ``Efficient
  signaling of spectral resources in spectrum pooling systems,'' in \emph{Proc.
  IEEE SCVT}, Eindhoven, The Netherlands, EU, Aug. 1--6, 2003.

\bibitem{li_icassp_2008}
F.~{Li} and J.~S. {Evans}, ``Optimal strategies for distributed detection over
  multiaccess channels,'' in \emph{Proc. IEEE ICASSP}, Las Vegas, NV, USA, Mar.
  30~--~Apr. 4, 2008.

\bibitem{zhang_twc_2009}
S.~{Zhang}, T.~{Wu}, and V.~K.~N. {Lau}, ``A low-overhead energy detection
  based cooperative sensing protocol for cognitive radio systems,''
  \emph{{IEEE} Trans. Wireless Commun.}, vol.~8, no.~11, pp. 5575--5581, Nov.
  2009.

\bibitem{yuan_dyspan_2007}
Y.~{Yuan}, P.~{Bahl}, R.~{Chandra}, P.~A. {Chou}, J.~I. {Ferrell},
  T.~{Moscibroda}, S.~{Narlanka}, and Y.~{Wu}, ``{KNOWS}: Kognitiv networking
  over white spaces,'' in \emph{Proc. IEEE DySPAN}, Dublin, Ireland, EU, Apr.
  17--20, 2007.

\bibitem{ma_dyspan_2005}
L.~{Ma}, X.~{Han}, and C.-C. {Shen}, ``Dynamic open spectrum sharing {MAC}
  protocol for wireless ad hoc networks,'' in \emph{Proc. IEEE DySPAN},
  Baltimore, MA, USA, Nov. 8--11, 2005.

\bibitem{wendong_commag_2007}
W.~{Hu}, D.~{Willkomm}, G.~{Vlantis}, M.~{Gerla}, and A.~{Wolisz}, ``Dynamic
  frequency hopping communities for efficient {IEEE} 802.22 operation,''
  \emph{{IEEE} Commun. Mag.}, vol.~45, no.~5, pp. 80--87, May 2007.

\bibitem{zhang_icc_2008}
W.~{Zhang}, K.~{Malik}, and K.~B. {Letaief}, ``Cooperative spectrum sensing
  optimization in cognitive radio networks,'' in \emph{Proc. IEEE ICC},
  Beijing, China, May 19--23, 2008.

\bibitem{pawelczak_wcm_2009}
P.~{Pawe{\l}czak}, S.~{Pollin}, H.-S.~W. {So}, A.~{Bahai}, R.~V. {Prasad}, and
  R.~{Hekmat}, ``Quality of service of opportunistic spectrum access: A medium
  access control approach,'' \emph{{IEEE} Wireless Commun. Mag.}, vol.~15,
  no.~5, pp. 20--29, Oct. 2008.

\bibitem{wellens_phycom_2009}
M.~{Wellens}, J.~{Riihij{\"a}rvi}, and P.~{M{\"a}h{\"o}nen}, ``Empirical time
  and frequency domain models of spectrum use,'' \emph{Elsevier Physical
  Communication Journal}, vol.~2, no. 1--2, pp. 10--32, Mar.--Jun. 2009.

\end{thebibliography}
\end{document}